\documentclass[twocolumn,prb,showpacs]{revtex4} %galley
\usepackage{graphicx}

\newcommand{\crossprod}{\times}
\newcommand{\be}{\begin{equation}}
\newcommand{\ee}{\end{equation}}
\newcommand{\barr}{\begin{eqnarray}}
\newcommand{\earr}{\end{eqnarray}}
\newcommand{\breakeq}{\nonumber \\ &&}
\newcommand{\mgrader}{^\circ} % degrees in math and equations.
\newcommand{\Tr}{\mathrm{Tr}} 
\newcommand{\Real}{\mathrm{Re}} 
\newcommand{\Imag}{\mathrm{Im}} 

\newcommand{\unit}[1]{\mathrm {\,#1}}

\begin{document}
%\draft

\title{Surface-enhanced Raman scattering and fluorescence near metal
nanoparticles}

\author{Peter Johansson}

\affiliation{Department of Natural Sciences, University of \"Orebro, 
   S-701 82 \"Orebro, Sweden}

\author{Hongxing Xu}

\affiliation{Division of Solid State Physics, 
   Lund University, Box 118, S--221\,00, Sweden}

\author{Mikael K\"all}

\affiliation{Department of Applied Physics, 
   Chalmers University of Technology, S--412\,96, G\"oteborg, Sweden} 

\date{\today}

\begin{abstract}
We present a general model study of surface-enhanced resonant Raman
scattering and fluorescence, focusing on the interplay between
electromagnetic (EM) effects and the molecular dynamics as treated by a
density matrix calculation.  The model molecule has two electronic
levels, is affected by radiative and non-radiative damping mechanisms,
and a Franck-Condon mechanism yields electron-vibration coupling.
The coupling between the molecule and the electromagnetic field is
enhanced by placing it between two Ag nanoparticles.
The results show that the Raman scattering cross section can, for
realistic parameter values, increase by some 10 orders of magnitude  (to
$\sim10^{-14}$ cm$^2$) compared with the free-space case.  Also the
fluorescence cross section grows with increasing EM enhancement,
however, at a slower rate, and this increase eventually stalls when
non-radiative decay processes become important.  Finally, we find that
anti-Stokes Raman scattering is possible with strong incident laser
intensities, $\sim 1$ mW/$\mu$m.
\end{abstract}

%PACS

%  78.30.-j, % Infrared and Raman spectra (for vibrational states in
%	     % crystals and disordered systems, see 63.20.-e and 63.50.+x
%	     % respectively)
%	     %
%  42.25.Fx  % Diffraction and scattering

\maketitle
%%%%%%%%%%%%%%%%%%%%%%%%%%%%%%%%%%%%

% BEGIN_MUL_COMM (sed tag)
%\begin{multicols}{2}
% END_MUL_COMM (sed tag)

\section{Introduction}
\label{sec:introduction}

% History etc. blah blah...
Surface-enhanced Raman scattering (SERS) was  discovered three decades
ago. SERS attracted a lot of attention through the
mid-1980's.\cite{Review1,Review2,SPIE}  During this period the basic
mechanisms behind the effect were studied, explained, and debated.  The
advent of single-molecule (SM) SERS\cite{Nie:97,Kneipp:97,Xu:99,Brus:99}
started a second period of intense research.  Now probably the prospect
of utilizing SERS and related spectroscopic techniques as an extremely
sensitive analytic tool, possibly combined with scanning probe
techniques,\cite{TERS} in a variety of life-science
applications\cite{Lakowicz:01} provides the main motivation for research
in the field.  But SM SERS experiments performed with intense lasers has
also raised new questions about the fundamental mechanisms
involved.\cite{Kneipp:96,Kneipp:98,Haslett:00,Brolo:04} 

It is generally agreed that electromagnetic (EM) enhancement effects are
the most important reason for the dramatic increase of the Raman
scattering cross-section $\sigma_R$ seen in SERS
experiments.
\cite{Mosk:78,Gersten:80,Takemori:87,GV:96,Xu:00,Corni:02,Stockman:03} 
In addition to this, $\sigma_R$ may also be enhanced due to charge transfer
effects.\cite{Persson:81,Pettinger:86,Otto:89,Brus:99,Otto:02}  
The EM enhancement effects received a lot of
attention from the theory side in the early days of SERS and has
continued to do so until now.  The electromagnetic enhancement has
in the general case not one single reason, but involves several, more or
less closely related aspects such as plasmon resonances, lightning rod
effects, and the formation of ``hot-spots'' in fractal clusters.

However, single-molecule Raman scattering is not possible without a
molecule that scatters the laser light inelastically.  In this paper we
focus on the interplay between the EM enhancement and molecular
dynamics, a topic that has received relatively limited attention in the
literature so far. A brief account of this work was published in Ref.\
\onlinecite{Xu:04}. We treat the molecule as an electronic two-level
system. Thus, we make no attempt of calculating {\it ab initio}
molecular properties, and charge transfer processes are also outside the
scope of this work. The focus is instead on calculating how the
molecule's different states are populated and how its coherent dipole
moment develops given a certain energy-level structure,
electron-vibration coupling, electromagnetic enhancement and laser
intensity. By treating the molecule dynamics within a density-matrix
calculation, we can evaluate a combined fluorescence and Raman spectrum
where also effects of electromagnetic enhancement are directly included.
Density-matrix methods were employed by Shen\cite{YRShen:74} to
distinguish between Raman scattering and hot luminescence, but to the
best of our knowledge they have not been used in the context of SERS.

% Old stuff
The results show how, for a molecule placed between two metallic
nanoparticles, both the fluorescence and in particular the Raman cross
sections are much larger than for a molecule in free space. This can be
discussed in terms of two EM enhancement factors, $M$ and $|M_d|$. Given an
electric-field enhancement $M(\omega)$ at the position of the molecule,
the Raman cross section increases by a factor $\sim |M|^4$ compared with
in free space, whereas the fluorescence cross section increases by the
factor $\sim |M|^4/|M_d|^2$. $|M_d|^2$ is a measure of how much the
decay rate of an excited state of the molecule is amplified near one or
several metal particles. For moderate to large molecule-particle
distances $M$ and $|M_d|$ are almost equal, but when the molecule gets
very close to a particle (a few nm or less) $|M_d|$ can be much larger
than $M$. Thus, surface enhancement of fluorescence is much less marked
than that of Raman scattering, and for small enough molecule-particle
separations the fluorescence cross section saturates.  When studying a
molecule next to a single metal particle, $M$ does not reach at all as
high values  as in the two-particle case, but $|M_d|$ still does.
Consequently, the SERS effect is largely absent in this case, and
fluorescence is strongly quenched when the molecule is placed close to
the particle. Distancing the molecule from the particle, $|M_d|$ drops and 
the fluorescence goes through a maximum and then eventually falls back
to the free-molecule value.

We have also studied how the Raman spectrum develops when the intensity
of the driving laser field is increased to relatively large values of
order $1 \unit{mW/(\mu m^2)}$.  In this case we find that anti-Stokes
Raman scattering becomes possible even if the molecule vibrations are
not thermally excited.  Such effects have been observed in a number of
SERS experiments, but the issue has been quite
controversial.\cite{Kneipp:96,Kneipp:98,Haslett:00} We find that
anti-Stokes Raman scattering in our model occurs because a vibrationally
excited level in the electronic ground state is populated by (possibly
repeated) excitation from the laser and subsequent deexcitation.  We
also find that when the laser intensity is increased to values where
Rabi oscillations become important, the anti-Stokes Raman cross section
decreases, and the corresponding peak in the spectrum broadens. 

The rest of the paper is organized as follows.  In Sec.\ \ref{sec:model}
we describe the model for the molecule that we use. As already stated,
the electromagnetic enhancement plays an important role in SERS, and
Sec.\ \ref{sec:em} describes how the EM enhancement of the incident
laser field, the emitted light, and the deexcitation of the
molecule is calculated.  Moreover, we also present numerical results for
some representative cases. Then in Sec.\ \ref{sec:free} we go back to
the model molecule and investigate how the parameter values entering the
model affect the absorption and Raman scattering cross sections for the
molecule {\em in free space}.  In Sec.\ \ref{sec:cross} we use
density-matrix theory to derive a general expression from which the
combined fluorescence and Raman scattering spectrum can be calculated
for a molecule placed near metal nanoparticles, and driven by, in
principle, an arbitrarily strong laser field. The so calculated spectra
are presented in Sec.\ \ref{sec:results}, and Sec.\ \ref{sec:summary}
concludes the paper with a brief comparison with experiments.  Finally,
Appendix \ref{app:em} gives some background information about the
electromagnetic calculations, and Appendix \ref{app:eh} presents a
derivation of the non-radiative damping of the molecule due to
electron-hole pair creation in the metal particles.

\section{Model}
\label{sec:model}

\begin{figure}[tb]
   \includegraphics[angle=0,width=8.0 cm]{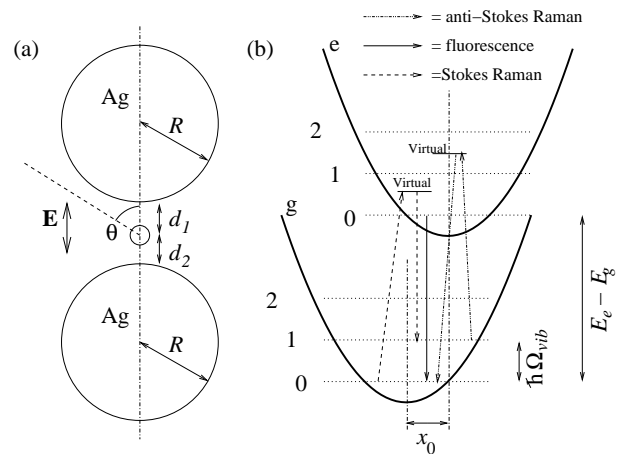}
   \caption{
   Schematic illustration of the model we use. (a) Electromagnetic
   enhancement is achieved by (usually) placing the model molecule
   between two metallic nanoparticles.
   (b) 
   The molecule has two electronic states (sometimes called bands in the
   following) each with a number of vibrational sublevels.
   The equilibrium position for the vibrational coordinate is displaced
   by $x_0$ upon electronic excitation; this mechanism provides
   electron-vibration coupling.  The arrows indicate typical pathways
   for the molecule state in fluorescence and Raman processes.
   }
   \label{fig:schematic}
\end{figure}

A schematic illustration of our model, both in terms of
molecule-nanoparticle geometry and the main features of the molecule
model is given in Fig.\ \ref{fig:schematic}.  In this figure we also
indicate the main steps involved in different Raman and fluorescence
processes.

We model the electronic degrees of freedom of the molecule as a
two-level system with states $|g\rangle$ and $|e\rangle$.  The frequency
$\Omega_{ge}$, defined from the energy-level separation $E_e-E_g$ as
\begin{equation}
   \hbar \Omega_{ge} = E_e-E_g,
\end{equation}
and the dipole matrix element, expressed as the product of the
elementary charge $e_e$ and a dipole length $\ell_{\mathrm{dip}}$,
$p_0=e_e\ell_{\mathrm{dip}}$, (to be further discussed below) are the
only parameters we need in this context to characterize the essential
electronic properties of the molecule.

To deal with Raman scattering we must of course introduce vibrational
degrees of freedom and an electron-vibration coupling.
We consider only one symmetric, vibrational mode, and let
$Q$ denote the corresponding coordinate.  The vibrational mode is
characterized by its angular frequency $\Omega_{\mathrm{vib}}$ and
reduced mass $\mu$.  The purely vibrational Hamiltonian can thus be
written
\begin{equation}
   H_{\mathrm{vib}} = \hbar \Omega_{\mathrm{vib}} (b^{\dag}b + \frac{1}{2}),
\end{equation}
where $b$ and $b^{\dag}$ are annihilation and creation
operators for a vibrational quantum. We assume that the vibrational
frequency is independent of the electron state, but the equilibrium
position is displaced a distance $x_0$ in the excited state.
Moreover, the transition dipole moment
introduced above is actually a function of $Q$: in the Born-Oppenheimer
approximation we get the Taylor expansion
$$
   p\equiv p(Q) = \langle e(Q) | e_e z | g(Q) \rangle 
   = p_0 + \left(\frac{dp}{dQ}\right) Q+\ldots,
$$
around $Q=0$. (Thus, we focus on the $z$ components of the electric
fields and dipole moments, since they are the ones enhanced in the
nanoparticle geometry we study.) Then the dipole matrix element between two
electron-vibration product states yields
\begin{eqnarray}
   &&
   \langle x_0;m | \langle e | e_e z | g  \rangle  0;n \rangle
   \breakeq
   =
   p_0 \, \langle x_0;m | 0;n \rangle
   +
   \left(\frac{dp}{dQ}\right) \, \langle x_0;m |Q| 0;n \rangle
   + \ldots,
\label{eq:dipseries}
\end{eqnarray}
where $|x_0;n\rangle$ denotes vibrational state $n$ in an oscillator
potential with the  equilibrium position displaced to $Q=x_0$, etc.  We
adopt the Condon approximation, thus retaining only the first (Albrecht
$A$) term\cite{Lombardi:86} in Eq.\ (\ref{eq:dipseries}). It gives the
dominating contribution in situations where the incident light is close
to resonance with the electronic transition.  We thus get
\begin{equation}
   \langle x_0; m | \langle e | e_e z | g  \rangle 0; n \rangle
   =
   p_0 f(n,m),
\end{equation}
where the Franck-Condon factor
\begin{eqnarray}
   &&
   f(n,m) 
   =\,  \langle 0;n|x_0;m \rangle
   =\,  \langle x_0;m|0;n \rangle
   \breakeq
   =
   \sqrt{n!m!}\,\, e^{-\alpha^2/2} 
   \sum_{k=0}^{min(n,m)} 
   \frac{(-1)^{(m+k)} \alpha^{n+m-2k} }{k!(n-k)!(m-k)!},
\end{eqnarray}
and the ratio  $\alpha$ between $x_0$ and the average zero-point vibration,
$(\langle (2Q)^2 \rangle)^{1/2}=
(2\hbar/(\mu \Omega_{\mathrm{vib}}))^{1/2}$,
\begin{equation}
   \alpha= x_0 \, \sqrt{\frac{\mu\Omega_{\mathrm{vib}}}{2\hbar}},
\end{equation}
serves as a measure of the electron-vibration coupling in the model.

The molecule is furthermore interacting with the electro\-magnetic field,
both with the incident laser field and the electromagnetic vacuum
fluctuations that cause spontaneous emission. 
The electric field originating from the laser can be written
\begin{equation}
   \vec{E}_L = 
   \hat{z} E_0 \cos{\Omega_L t} = 
   \hat{z} ({E_0}/{2})\, [e^{i\Omega_L t} + e^{-i\Omega_L t}],
\label{eq:laserfield}
\end{equation}
so that the nominal incident intensity is 
$
   I_{\mathrm{in}} = c\varepsilon_0 E_0^2/2.
$
We model the laser-molecule interaction within the dipole approximation
by a 
$-e_{e}\, \vec{r}\cdot\vec{E}$ term, 
$H'=-e_{e}E_0z [e^{i\Omega_L t} + e^{-i\Omega_L t}]/2$ in the Hamiltonian.
When evaluating matrix elements due to $H'$ we adopt
the rotating wave approximation (RWA). Only the part of the field
varying with time as $e^{-i\Omega_L t}$ is kept in the matrix
elements when the molecule is excited, and vice versa when it is
deexcited.  This yields 
\begin{equation}
   \langle x_0;m | \langle e | H' | g\rangle |0;n \rangle
   =
   -(p_0E_0/2) e^{-i\Omega_L t} \, f(n,m),
\end{equation}
and 
\begin{equation}
   \langle 0;n | \langle g | H' | e \rangle |x_0;m \rangle
   =
   -(p_0E_0/2) e^{i\Omega_L t} \, f(n,m).
\end{equation}
(In the following, when the local field at the molecule is modified by
the nearby metallic particles, these matrix elements must be corrected
by an enhancement factor.)

The interaction between the molecule and the
vacuum fluctuations of the electromagnetic field can be described by a
Hamiltonian 
$
   H_{\mathrm{fluct}} = 
   -e_{e} z  E_{\mathrm{vac}} \cdot \hat{z},
$
where the corresponding electric field must be given on second-quantized
form.
In free space we have
\begin{equation}
   \vec{E}_{\mathrm{vac}}(\vec{r},t)
   =
   \sum_{{\bf k},\alpha}
   \sqrt{\frac{\hbar}{2\epsilon\omega_{k}}} 
   \frac{i \omega_{k}}{L^{3/2}}
   {\bf \varepsilon}_{{\bf k},\alpha}
   e^{i{\bf k}\cdot{\bf r}}
   [a_{{\bf k},\alpha}(t)-a_{{\bf k},\alpha}^{\dag}(t)],
\end{equation}
where the electromagnetic field has been normalized in a box with side
$L$, and 
$a_{{\bf k},\alpha}$ and $a_{{\bf k},\alpha}^{\dag}$ are
annihilation and creation operators for photons. 
For a molecule in free space the interaction with the vacuum
fluctuations gives a transition rate
\begin{equation}
   \gamma_{\mathrm{rad},0} (n,m) = 
   \frac{\omega^3}{3\pi \hbar \varepsilon_0 c^3} |p_0|^2 |f(n,m)|^2,
\label{eq:gammarad0}
\end{equation}
where $\omega$ is the angular frequency corresponding to the transition
energy, i.e.\
\begin{equation}
   \omega = 
   \Omega_{ge} 
   + (m-n) \omega_{\mathrm{vib}}.
\end{equation}
With a dipole moment corresponding to $\ell_{\mathrm{dip}}=1$ \r{A}, a
transition energy of 2.5 eV, and $|f|^2=1$, we get numerically
$\gamma_{\mathrm{rad},0}\approx 5.9\times10^{7} \unit{s^{-1}}$.

This transition rate is enhanced when the molecule is placed near
metallic nanoparticles; the radiative losses increase and in addition
energy can be dissipated in the particles, thus 
$$
   \gamma_{\mathrm{rad},0} (n,m)  
   \rightarrow
   |M_d(\omega)|^2
   \gamma_{\mathrm{rad},0} (n,m). 
$$
We will discuss the dissipation enhancement factor $|M_d(\omega)|^2$
further in Sec.\ \ref{sec:em}.

Our model also includes, at a phenomenological level, a few more
relaxation mechanisms. Vibrational damping is described by the parameters
$\gamma_{\mathrm{vib}}^{(g)}$ and 
$\gamma_{\mathrm{vib}}^{(e)}$. The transition rate from a state with $n$
vibrational quanta to the one with $n-1$ is given by 
$n \gamma_{\mathrm{vib}}^{(g)}$ and 
$n \gamma_{\mathrm{vib}}^{(e)}$, in the electronic ground and
excited states, respectively. The electronic state of the molecule does
not change in this process.
Finally, we include a dephasing rate $\gamma_{ph}$ affecting the
coherent dipole moment of the molecule.
The primary effect of the dephasing rate on the calculated results
is to broaden the fluorescence resonances of the molecule.
In reality an organic molecule has many vibration modes, and therefore
an almost continuous fluorescence spectrum. Dephasing gives us a
broadened fluorescence spectrum even though the model molecule has only
one vibrational mode. Similar broadening parameters are also used in
Raman cross section calculations that account for molecular structure in
more detail.\cite{Kelley:03}

\section{Electromagnetic enhancement}
\label{sec:em}

\subsection{Theory}

In this section we address the calculation of the electromagnetic enhancement 
factors $M$ and $|M_d|$ introduced above.
We first assume that the  system of metal particles and the molecule is
illuminated by a $p$ polarized plane wave with angular frequency
$\omega$ and wave number $k=\omega/c$ incident from a direction
specified by the angles $\theta$ and $\varphi$.
The corresponding incident electric field can be written 
\begin{equation}
   \vec{E}_{\mathrm{in}} (\vec{r})
   =
   \vec{E}_{0}
   e^{i (\vec{k}\cdot\vec{r}-\omega t)},
\label{eq:planewave}
\end{equation}
with 
\begin{equation}
   \vec{E}_{0} 
   = 
   E_0 
   (
      \hat{x}\cos{\theta}\cos{\varphi}
      +
      \hat{y}\cos{\theta}\sin{\varphi}
      -
      \hat{z}\sin{\theta}
   ),
\end{equation}
(in the following we assume the time-dependence of all fields to be
$e^{-i \omega t}$).
When this field impinges on the metal particles they are 
polarized, plasmons may be excited, and, most importantly, the
electromagnetic interaction between the spheres will cause a field
enhancement in the region of space in between them.  We
are mainly interested in calculating the $z$ component of the electric field
at the position of the molecule ($\vec{r}=\vec{0}$); it is this field that
excites the dipole moment of the molecule when a laser beam illuminates
the system. 
We define the enhancement factor $M(\omega)$ in terms of the induced
total field at the position of the molecule, through the relation
\begin{equation}
   M(\omega,\theta)    
   =
   {\hat{z}\cdot \vec{E}_{\mathrm{tot}} (\vec{0})}
   /
   {|\vec{E}_0|}.
\end{equation}
The enhancement of the incident laser field is thus given by
$M(\Omega_L,\theta_{\mathrm{in}})$.  Moreover, as a consequence of
electromagnetic reciprocity, the field sent out in the direction of
$\theta$ and $\varphi$ by an oscillating 
dipole (angular frequency $\omega$) placed at
$\vec{r}=\vec{0}$ is equally enhanced by a factor $M(\omega,\theta)$.

% Mie ansatz
To calculate the local field and $M$ we must find the electric field
around the nanoparticles. To this end we employ extended Mie theory,
expanding the field around each of the two spheres in terms of vector
spherical harmonics representing magnetic and electric
multipoles,\cite{Jackson,Waterman}  
\begin{equation}
   \vec{E} (\vec{r})
   =
   \sum_{\tau,\ell,m}
   \left[
      a_{\tau\ell m}^{s}
      \vec{\psi}_{\tau\ell m}^{(R)}(\vec{r}-\vec{r}_s)
      +
      b_{\tau\ell m}^{s}
      \vec{\psi}_{\tau\ell m}^{(O)}(\vec{r}-\vec{r}_s)
   \right].
\label{eq:Mieansatz}
\end{equation}
Here the functions $\vec{\psi}^{(R)}$ and $\vec{\psi}^{(O)}$ represent
waves that are regular at the origin and outgoing from the sphere,
respectively.
The index $s$ tells whether we are referring to the upper ($s=1$)
sphere, centered at $\vec{r}_1$, or the lower one ($s=2$). 
The rest of the indices indicate the values of the
angular momentum of the multipole $\ell$ and $m$, and whether it is a
magnetic ($\tau=1$) or electric multipole ($\tau=2$). 
The corresponding basis
functions are given in terms of vector spherical harmonics
$\vec{X}_{\ell m}$,\cite{Jackson}
\be
   \vec{\psi}_{1 \ell m}
   =
   \vec{X}_{lm}(\theta,\varphi)
   z_l(kr)
\ee
and 
\be 
   \vec{\psi}_{2\ell m}
   =
   k^{-1}
   \nabla\crossprod
   \left\{
      \vec{X}_{lm}(\theta,\varphi)
      z_l(kr)
   \right\},
\ee
where $z_l$ is either a spherical Bessel function $j_l$ for the regular
waves, or a spherical Hankel function $h_{\ell}$, for the outgoing
waves.  The expression in Eq.\ (\ref{eq:Mieansatz}) is valid for points
situated between sphere $s$ and a larger concentric sphere that just
touches the other sphere.  For a general point outside the two spheres,
on the other hand, the field must be written as a sum of the outgoing
waves generated by both the spheres, plus the incident wave driving the
system.  

% Sphere response
The $a$ and $b$ coefficients appearing in Eq.\ (\ref{eq:Mieansatz}) are
related  by sphere response functions $s_{1\ell}$ and $s_{2\ell}$
depending on the radius
$R$ of the sphere and its dielectric properties characterized by the
local dielectric function $\epsilon$ (taken from Ref.\
\onlinecite{Palik}) and wave number
$k_r=\sqrt{\epsilon}\omega/c$. We get
\begin{equation}
 s_{1\ell} = \frac{b_{1\ell}}{a_{1\ell}} =
 -
 \frac{ 
k R\, j_l'(k R) 
- 
j_l (k R)\, \mathcal{J}_l 
}
  { k R\, h_l'(k R) -  h_l (k R)\, \mathcal{J}_l },
\label{eq:slMres}
\end{equation}
for the magnetic multipoles, and
\begin{equation}
 s_{2\ell} = \frac{b_{2\ell}}{a_{2\ell}} =
 -\, \frac{\epsilon\, k R\, j_l'(k R) 
          + j_l (k R)\, (\epsilon-1-\mathcal{J}_l) }
   {\epsilon\, k R\, h_l'(k R) 
          + h_l (k R)\, (\epsilon -1 -\mathcal{J}_l) },
\label{eq:slEres}
\end{equation}
for the electric multipoles. $\mathcal{J}_l$ is shorthand for
$$ \mathcal{J}_l= k_r R\, j_l'(k_rR)/j_l(k_rR). $$

In view of these relations between incident and outgoing waves, we have
full knowledge of the electromagnetic field once we know the $a$
coefficients on both spheres. We solve for them by realizing that the
waves incident on a sphere either originate from the external source or
from the waves scattered off the other sphere, and this yields
\begin{equation}
   a_{\tau\ell m}^{s} 
   =
   a_{\tau\ell m, \mathrm{ext}}^{s} 
   +
   \sum_{\tau' \ell'}
   \tilde{c}_{\tau'\ell'm,\tau\ell m}
      (\vec{r}_s - \vec{r}_{\overline{s}})
   \,
   s_{\tau'\ell'}\,a_{\tau'\ell' m}^{\overline{s}}.
\label{eq:asystem}
\end{equation}
Here $\overline{s}$  stands for the {\em other} sphere (i.e.\
$\overline{1}=2$ etc.).   Expressions for the $\tilde{c}$ and
$a_{\mathrm{ext}}$ coefficients can be found in Appendix \ref{app:em}.

After solving the system of equations obtained from (\ref{eq:asystem}), also
the $b$ coefficients can be determined thanks to Eqs.\ (\ref{eq:slMres})
and (\ref{eq:slEres}). Then, as discussed above, the electric field can
be calculated anywhere in space, in particular the local field at the
molecule is given by 
\begin{equation}
   \vec{E}(\vec{0})
   =
   \vec{E}_{0}
   +
   \sum_{\tau,\ell, m, s}
   b_{\tau\ell m}^{s} \vec{\psi}_{\tau\ell m}(-\vec{r}_s),
\end{equation}
from which we can find the enhancement factor $M$.
It should be noted that in the calculation of the $z$ component of the
electric field on the symmetry axis there are only contributions from
$m=0$ terms.

% How to get M_d.
When calculating $|M_d|$, we cannot use the plane wave as a source,
instead we place an oscillating point dipole at the position of the
molecule,
\begin{equation}
   \vec{p}(t) 
   =
   \Real
   \left[
      p e^{-i\omega t}
   \right].
\end{equation}
In free space, it would send out radiation with a total (time-averaged)
power,\cite{Jackson}
\begin{equation}
   P_{\mathrm{free}} = \frac{\omega^4 p^2}{4\pi \varepsilon_0 \, 3c^3}.
\end{equation}
With the spheres present we can once again solve for the total electric
field from Eq.\ (\ref{eq:asystem}). The only difference 
is that now $a_{\mathrm{ext}}$ appropriate for a localized
dipole source has to be used, see Appendix \ref{app:em}.
% fill in a formula here
Once the field has been calculated we integrate the time-averaged
Poynting vector over a small sphere enclosing the dipole to find the
total radiated power $P_{\mathrm{Mie}}$, which yields a contribution to
the damping enhancement rate $|M_d|^2$,
\begin{equation}
   |M_d(\omega)|^2 
   =
   {P_{\mathrm{Mie}}}/P_{\mathrm{free}}
\end{equation}
$P_{\mathrm{Mie}}$ accounts for losses of energy due to radiation
leaving the molecule in all directions, as well as for dielectric losses
in the particles, i.e.\ energy from the molecule  that goes to heating
the metallic particles.

In practice we need to add an another contribution to $|M_d|^2$ that is
due to electron-hole pair creation in the metal particles when the
molecule is placed very close to them.\cite{Liebsch:87,Barnes:98,Larkin:04}  
This involves processes that depend on the non-local optical response of
the metal particles and which are therefore not included in the
calculation discussed above. The dielectric losses captured by the Mie
calculation at short molecule-particle separations  $d$ scale as $1/d^3$
as a result of the distance dependence of the molecule's dipole field.
The power loss caused by electron-hole pair creation, on the other hand,
scales as $1/d^4$, and becomes a dominant damping process for small $d$
($d$ less than 1 nm or so).  In Appendix \ref{app:eh} we give a detailed
presentation of the calculation of the power loss $P_{eh}$ due to
electron-hole pair creation in the metal particles.  After $P_{eh}$ is
calculated the total damping rate enhancement is found as 
\begin{equation}
   |M_d(\omega)|^2 = (P_{\mathrm{Mie}} + P_{eh})/ P_{\mathrm{free}}.
\label{eq:Mdtot}
\end{equation}

\subsection{Results for the enhancement}

\begin{figure}[tb]
   \includegraphics[angle=0,width=7.0 cm]{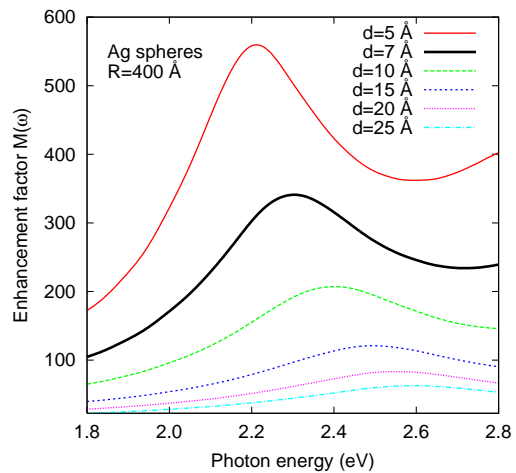}
   \caption{(color online)
   Calculated enhancement factors $M(\omega)$ for the incident light as
   a function of photon energy, for a molecule symmetrically  placed
   between two Ag spheres with 400 \r{A} radius.  The different curves
   give results for a number of different molecule-particle separations
   $d$.
   }
   \label{fig:Mfreq}
\end{figure}

\begin{figure}[tb]
   \includegraphics[angle=0,width=7.0 cm]{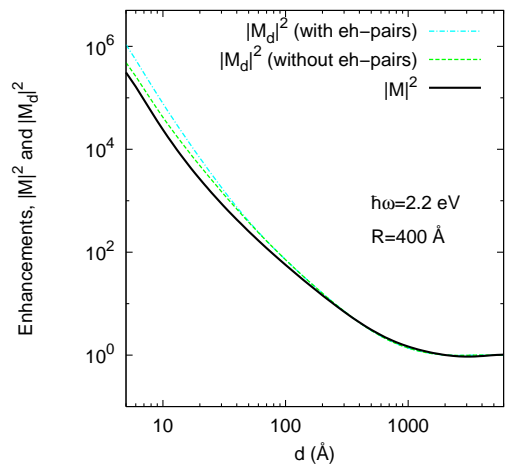}
   \caption{(color online)
      Calculated enhancement factors $|M|^2$ (full curves) and $|M_d|^2$
      as a function of the molecule-particle separation $d$. As in Fig.\
      \ref{fig:Mfreq},  the molecule is symmetrically placed between two
      Ag spheres of radius 400 \r{A}, and the photon energy is 2.2 eV\@.
      The dashed curves give $|M_d|^2$ {\em without} the contribution
      due to electron-hole pair creation, whereas the dash-dotted curves
      show $|M_d|^2$ {\em with} these contributions included.
   }
   \label{fig:Md}
\end{figure}

\begin{figure}[tb]
   \includegraphics[angle=0,width=7.0 cm]{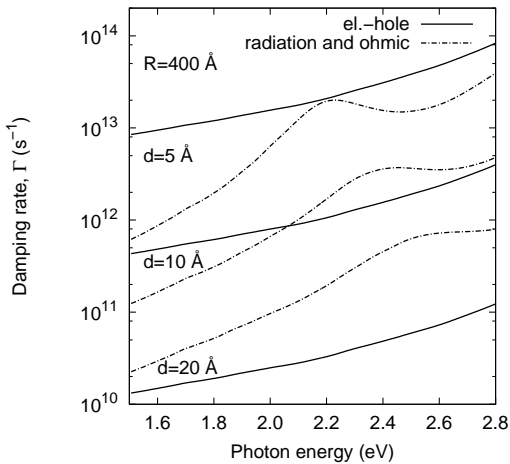}
   \caption{
      Calculated damping rates as a function of photon energy. As in
      Figs.\ \ref{fig:Mfreq} and \ref{fig:Md} the molecule is
      symmetrically placed between two 400 \r{A} radius silver spheres.
      The curves show results for three different molecule-particle
      separations $d$ for the separate contributions from, on one hand,
      electron-hole pair creation and, on the other hand, losses
      captured by the Mie calculations (here labeled as radiation and
      ohmic damping). The damping rate for the free molecule was
      calculated from Eq.\ (\ref{eq:gammarad0}) with
      $\ell_{\mathrm{dip}}=1$ \r{A}.
   }
   \label{fig:damprate}
\end{figure}

Figures \ref{fig:Mfreq}, \ref{fig:Md}, and \ref{fig:damprate} show 
calculated results related to the electromagnetic enhancement. In these
calculations the molecular response which  we address in more detail
later 
plays no role, however, the placement of the molecule, i.e. where to evaluate
the electric fields, is crucial. 
We consider a case where the molecule is placed
symmetrically between the two metal nanoparticles, $d$ is the
molecule-particle separation, and consequently the smallest distance
between the spheres is $2d$. 

Figure \ref{fig:Mfreq} shows results for  $M(\omega, \theta=90\mgrader)$
as a function of photon energy for a few different values of $d$.
There is an overall increase of the enhancement when the
spheres approach each other. Moreover, the enhancement
factor has a resonance peak with a resonance frequency that shifts with
$d$. For $d=25$ \r{A} it lies somewhat above 2.6 eV, but it redshifts
with decreasing $d$ and for $d=5$ \r{A} ends up at $\approx$ 2.2 eV.
The resonance is caused by the interaction between the plasmon modes of 
the two  separate spheres leading to the formation of a coupled mode
with surface charges of opposite sign facing each other across the gap
between the spheres.  This means that a smaller $d$ gives a more
charge-neutral mode, and thereby smaller restoring forces leading to the
redshift shown in Fig.\ \ref{fig:Mfreq}.

In Fig.\ \ref{fig:Md} we have plotted the enhancement factors
$|M(\omega)|^2$ and $|M_d(\omega)|^2$ (for the latter quantity both with
and without contributions due to electron-hole pair creation) as a
function of the molecule-particle separation $d$ for photon energy
$\hbar\omega=2.2 \unit{eV}$. Quite naturally, since $d$ spans a large
range of values, so do the enhancement factors, from about 1 for $d\agt
1000 \unit{\r{A}}$, to $10^5$ or more at the smallest $d<10
\unit{\r{A}}$.  For distances $d$ larger than 200 \r{A} the
three curves follow each other very closely; at least on this scale no
difference is discernible. This is because the silver particles are
close enough to enhance the incident field and thereby, as a consequence
of electromagnetic reciprocity, also enhance the radiation rate and
$|M_d|$. But the particles are still sufficiently far away from each other
and the molecule that the dielectric losses are negligible. Thus,
radiation losses completely dominate other damping mechanisms here.
Continuing towards smaller $d$, the two $|M_d|^2$ curves separate
themselves from the $|M|^2$ curve because now losses in the silver
particles are no longer negligible compared with radiation losses. The
inclusion of damping due to electron-hole pair creation does not make
any difference at first; non-local effects play no role and all loss
mechanisms are already accounted for in the Mie calculation. It is only
when $d$ reaches values of $\approx 30 \unit{\r{A}}$ or below that the
two curves representing $|M_d|^2$ with and without electron-hole pair
damping begin to differ.

Figure \ref{fig:damprate} shows the separate contributions to the
damping rate from electron-hole pair creation and the remaining,
radiative and dielectric loss mechanisms. As in the previous figure we
see a rapid increase of the damping rate with decreasing $d$. This
tendency is more marked for the electron-hole pair losses which vary as
$1/d^4$.  The electron-hole pair losses constitute a minor contribution
for $d=20 \unit{\r{A}}$, but has at $d=5 \unit{\r{A}}$ become the
dominant damping mechanism.

\section{The molecule in free space}
\label{sec:free}

Given the model for the molecule discussed above, we can calculate the
absorption and Raman scattering cross-sections for a
molecule in free space using lowest-order perturbation theory, i.e.\ the
Fermi golden rule.\cite{Sakurai}

The absorption cross-section derived in this way reads,
\begin{equation}
   \sigma_A
   =
   p_0^2 \, 
   \frac{\Omega_L}{c \varepsilon_0}
   \sum_{n}
   \frac{ \hbar \Gamma_{\mathrm{tot}} \, f(0,n) f(0,n)}
   {(\hbar\Omega_L - n\hbar\omega_{\mathrm{vib}}
   -\hbar\Omega_{ge})^2 +\hbar^2  \Gamma_{\mathrm{tot}}^2}
\end{equation}
where
$\Gamma_{\mathrm{tot}} = \gamma_{ph}+ \gamma_{\mathrm{vib}}/2 +
\gamma_{\mathrm{rad},0}/2$
and $\gamma_{\mathrm{rad},0}$ and $\gamma_{\mathrm{vib}}$ are the
radiative and vibrational decay rates of the final, excited state. 
For the combinations of parameter values that we use,
$\gamma_{ph}$ gives the completely dominating contribution to
$\Gamma_{\mathrm{tot}}$ for a molecule in free space.

For the (fundamental) Raman cross-section we get in a similar way, using
the Fermi golden rule with a second-order transition matrix element,
\begin{equation}
   \sigma_{R}
   =
   p_0^4\,
   \frac{\omega'^3\, \Omega_L}{6 \pi \varepsilon_{0}^{2} c^4}
   \left|
      \sum_{n}
      \frac{f(1,n)\,f(0,n)}
      {\hbar\Omega_L - n\hbar\omega_{\mathrm{vib}} - \hbar\Omega_{ge} 
	 + i\hbar \Gamma_{\mathrm{tot}}}
   \right|^2.
\end{equation}
Here $\omega'=\Omega_L-\omega_{\mathrm{vib}}$, and 
$\Gamma_{\mathrm{tot}} = \gamma_{ph}+ \gamma_{\mathrm{vib}}/2 
+ \gamma_{\mathrm{rad},0}/2$
for the intermediate, virtual state.
 
% \begin{figure}[tb]
%    \includegraphics[angle=0,width=8.0 cm]{M_func_of_w.eps}
%    \caption{
%       Calculated enhancement factors $|M|^2$ (full curves) and $|M_d|^2$
%       as a function of photon energy, for a molecule symmetrically
%       placed between two Ag spheres of radius 400 \r{A}.  The
%       molecule-particle separation is 5 \r{A} for the upper three
%       curves, and 25 \r{A} for the lower three curves. The dashed curves
%       give $|M_d|^2$ {\em without} the contribution due to electron-hole
%       pair creation, whereas the dash-dotted curves show $|M_d|^2$ 
%       {\em with} these contributions included.
%    }
%    \label{fig:Mfreq}
% \end{figure}

% Heights outside 320, 200 and 80, 55
\begin{figure}[tb]
   \includegraphics[angle=0,width=8.0 cm]{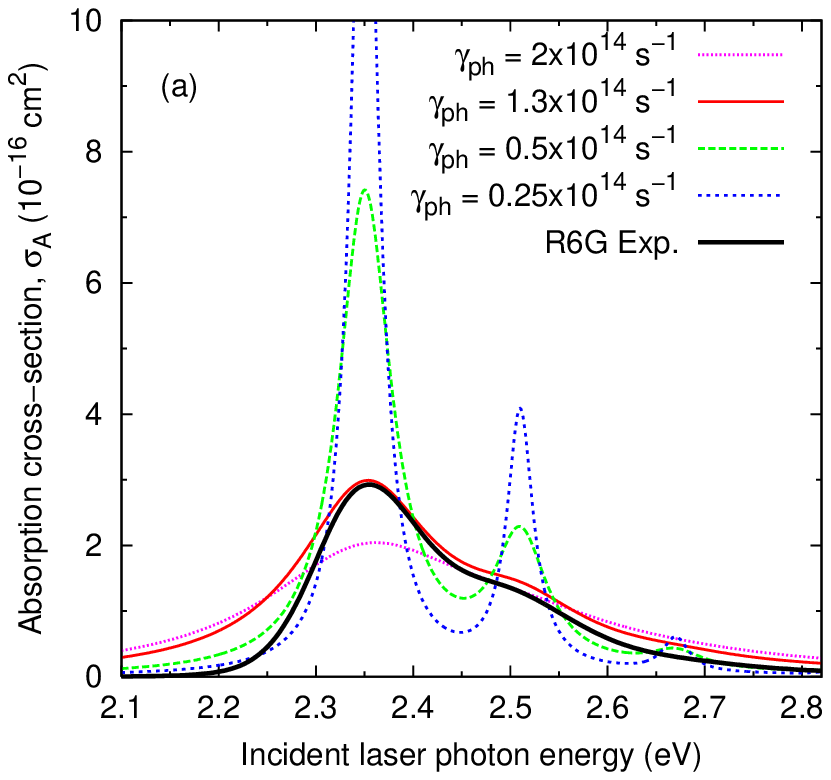}
   \includegraphics[angle=0,width=8.0 cm]{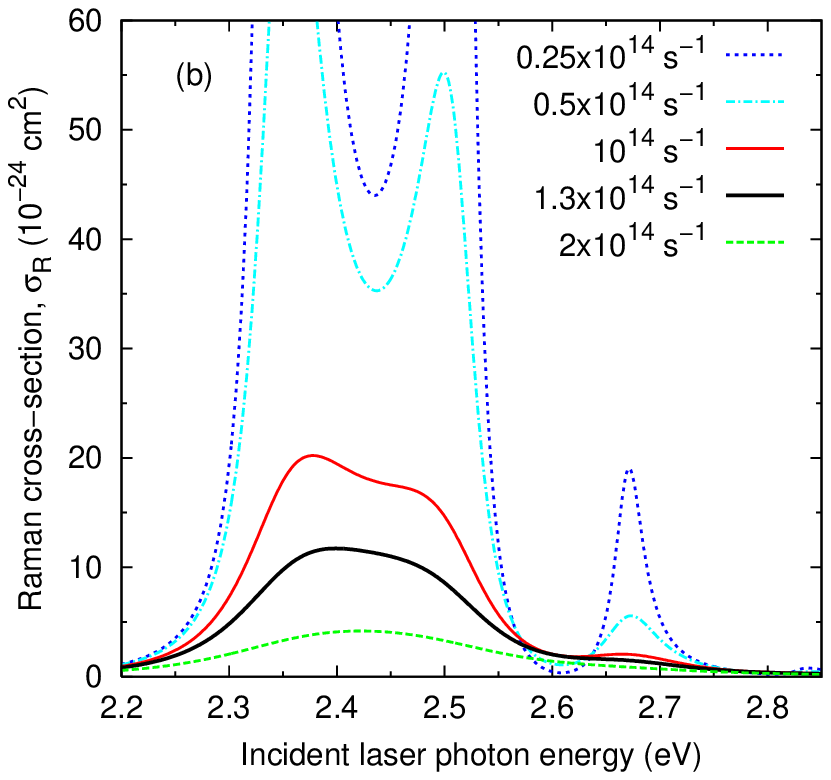}
   \caption{(color online)
      Calculated (a) absorption and (b) Raman  profiles (i.e.\ cross
      sections as a function of incident photon energy) for a molecule
      in free space for a number of
      different dephasing rates $\gamma_{ph}$.  The remaining parameter
      values used in the calculation are $\ell_{\mathrm{dip}}=1.2
      \unit{\r{A}}$, 
      $\hbar\Omega_{ge}= 2.35 \unit{eV}$,
      $\hbar\Omega_{\mathrm{vib}}=160 \unit{meV}$,
      and $\alpha=0.5$.
   }
   \label{fig:profiles}
\end{figure}

% Absorption and Raman profiles and parameter values.
% Do we have a reference to the R6G spectrum?
Turning to calculated results for spectra, we begin by looking at
absorption and Raman profiles and their dependence on the molecular
parameter values. Figure \ref{fig:profiles} shows absorption and Raman
profiles plotted for a series of different dephasing rates
$\gamma_{ph}$.  In panel (a) we also give experimental  results for the
absorption cross section of a commonly used fluorescent dye molecule,
Rhodamine 6G (R6G).  This comparison provides some indications on what
parameter values can be considered realistic. Thus we have chosen
$\hbar\Omega_{ge}=2.35 \unit{eV}$ to get about the same peak position in
the model calculation as for R6G. The value used for
$\hbar\Omega_{\mathrm{vib}}$, 160 meV,   is characteristic for a C-C
stretch vibration.
The dipole length
$\ell_{\mathrm{dip}}$ and the dephasing rate largely determine the
height and width of the absorption peak. We have set
$\ell_{\mathrm{dip}}=1.2 \unit{\r{A}}$, while we in the calculations
presented in Fig.\ \ref{fig:profiles} used a number of different values
for $\gamma_{ph}$ to study its effects.  Obviously
$\gamma_{ph}=1.3\times10^{14} \unit{s^{-1}}$ gives the best agreement 
with the R6G experimental spectrum; the remaining difference is the
broader wings (especially on the low-frequency side) of the calculated
spectrum.  Finally, the Franck-Condon
parameter $\alpha$ mainly determines the strength of absorption
sidebands (or shoulders) due to vibrational excitations. In the
calculations presented here we have used $\alpha=0.5$ and this gives a
shoulder in the absorption spectrum of similar strength to that found in
the R6G spectrum.

% Absorption profiles as a function of the dephasing parameter.
As can be seen in Fig.\ \ref{fig:profiles} (a), varying $\gamma_{ph}$
results in variations of the width and height of the absorption peaks;
the widths are proportional to $\gamma_{ph}$ whereas the heights are
approximately inversely proportional to $\gamma_{ph}$. Thus, with
$\gamma_{ph}$ considerably smaller than $1.3\times10^{14} \unit{s^{-1}}$
the vibrational sidebands create marked, resonant peaks at 2.51 eV and
2.67 eV, while with a larger $\gamma_{ph}$ the absorption spectrum is
more broadened.

% Raman profiles as a function of dephasing.
% Figure \ref{fig:profiles} (b) shows the Raman profiles (i.e.\ the Raman
% scattering cross-section as a funtion of the incident laser photon
% energy) for a number of different choices for $\gamma_{ph}$. As for the
% absorption profile, a small value for $\gamma_{ph}$ gives a profile with
% high sharp peaks. This is even more marked for the Raman profile at
% least for the range of laser frequencies close to resonance.  In this
% situation [in Fig.\ \ref{fig:profiles} (b) primarily between 2.3 eV and
% 2.4 eV] the width varies as $\sim \gamma_{ph}^2$.  
% On the other hand, if the laser frequency is further away from
% resonance, the detuning not the dephasing rate dominates the eenrgy
% denominator entering the Raman scattering matrix element and in this
% case $\sigma_R$ is less dependent on $\gamma_{ph}$.

% Raman profiles as a function of dephasing.
The Raman profiles displayed in Fig.\ \ref{fig:profiles} (b) show the
same trends as the absorption profiles when the dephasing rate
$\gamma_{ph}$ is varied, but the dependence is somewhat more complicated
in this case. The Raman scattering cross section is governed by a
second-order matrix element with an energy denominator with a real part
set by the laser detuning and an imaginary part mainly set by
$\gamma_{ph}$.  Thus in the range of laser frequencies close to
resonance we get peaks in the profile with a height determined by
$\gamma_{ph}$.  In this situation [in Fig.\ \ref{fig:profiles} (b)
primarily between 2.3 eV and 2.5 eV] the height varies as $\sim
1/\gamma_{ph}^{2}$ and the width as $\sim \gamma_{ph}^2$.
On the other hand, further away from resonance the detuning, not the
dephasing rate, dominates the energy denominator entering the Raman
scattering matrix element and in this case $\sigma_R$ varies much more
weakly with $\gamma_{ph}$.

\section{Cross-section calculation}
\label{sec:cross}

In this section we outline in detail the calculation of the spectrum of
light emitted by the molecule.  This includes both light scattered as a
result of Rayleigh or Raman processes, and fluorescence as a result of
electronic transitions in the molecule. The methods presented below
makes it possible to carry out this calculations in a general case with
strong enhancement of both the incident light and of the damping rate.

\subsection{Emitted intensity}

The emitted light intensity at the point ${\bf r}_0$ in the far field can
be written\cite{Scully} 
\begin{equation}
   I_{\mathrm{em}} 
   = 
   2 \varepsilon_0 c 
   \langle 
      E_{\theta}^{(-)}({\bf r}_0, t)\, 
      E_{\theta}^{(+)}({\bf r}_0, t)
   \rangle.
\label{eq:intensity1}
\end{equation}
Here $E_{\theta}^{(+)}({\bf r}_0,t)$ and $E_{\theta}^{(-)}({\bf r}_0,t)$
stand for the positive and negative frequency parts of the $\theta$
component of the electric field, respectively. By using the
normal-ordered correlation function 
$\langle E_{\theta}^{(-)}({\bf r}_0, t)\, 
   E_{\theta}^{(+)}({\bf r}_0, t)\rangle$
we are assured that the vacuum fluctuations do not contribute to
$I_{\mathrm{em}}$.

Our present goal is not only to find the total intensity of emitted light,
but also its spectral distribution.  Starting from Eq.\
(\ref{eq:intensity1}), and using the Wiener-Khintchine theorem this can
be written
\begin{equation}
   I_{\mathrm{em}}(\omega)
   =
   \frac{1}{\pi}
   \Real
   \int_{0}^{\infty}
   d \tau \,
   2 \varepsilon_0 c  \,
   \langle 
      E_{\theta}^{(-)}({\bf r}_0, 0)\, 
      E_{\theta}^{(+)}({\bf r}_0, \tau)
   \rangle
   \,
   e^{i\omega \tau}.
\label{eq:spectrum1}
\end{equation}
The electric fields are caused by the electric dipole
moment $p(t)$ of the molecule.
A point dipole with dipole moment $\hat{z}pe^{-i\omega t}$ placed at
the origin in free space generates the electric field
% \begin{equation}
%    \vec{E}
%    =
%    \hat{\theta} p \frac{\omega^2 \sin{\theta}}{4\pi\varepsilon_0 c^2 r}
%    e^{i(kr-\omega t)}
% \label{eq:Efromdip}
% \end{equation}
$
   \vec{E}
   =
   \hat{\theta} \, p \, 
   e^{i(kr-\omega t)} \,
   {\omega^2 \sin{\theta}}/{(4\pi\varepsilon_0 c^2 r)}
$
at a distance $r$ from the dipole.
If this expression is combined with Eq.\ (\ref{eq:spectrum1})
we get the differential scattering and fluorescence cross-section
\begin{equation}
   \frac{d^2\sigma}{d\Omega d(\hbar\omega)}
   =
   \frac{\omega^4 \sin^2{\theta}}{I_{in} 8 \pi^3 c^3 \varepsilon_0\hbar}
   \Real
   \int_{0}^{\infty}
   d \tau \,
   e^{i\omega \tau} \,
   \langle 
      p^{(-)}(0)\, 
      p^{(+)}(\tau)
   \rangle.
\label{eq:spectrum2}
\end{equation}

When the molecule is no longer placed in free space the electromagnetic
propagation from source to detector is modified. This can be
described by the enhancement factor $M(\omega,\theta)$ introduced in
Sec. \ref{sec:em}.  $M(\omega, \theta)$ is closely related to
a photon Green's function.\cite{Johansson:02}  The $\sin{\theta}$ factor
in Eq.\ (\ref{eq:spectrum2}) is absorbed in $M$ which yields the emitted
spectrum (in the numerical results presented below we use
$\theta=90\mgrader$ as the observation angle) from a molecule placed in
the vicinity of one or several metal nanoparticles,
\begin{equation}
   \frac{d^2\sigma}{d\Omega d(\hbar\omega)}
   =
   \frac{\omega^4 |M(\omega, \theta)|^2}
   {I_{in} 8 \pi^3 c^3 \varepsilon_0\hbar}
   \Real
   \int_{0}^{\infty}
   d \tau \,
   e^{i\omega \tau} \,
   \langle 
      p^{(-)}(0)\, 
      p^{(+)}(\tau)
   \rangle.
\label{eq:spectrum3}
\end{equation}
To get any further we must evaluate the dipole-dipole correlation
function $ \langle p^{(-)}(0)\, p^{(+)}(\tau) \rangle$ by solving the
equations of motion for the molecule dynamics.  We will deal with this
using density-matrix methods.

\subsection{Equation of motion for the density matrix}

We consider a model of the molecule with in total $N$ quantum states: two
electronic levels, $|g\rangle$ and $|e\rangle$, sometimes
called bands in the following, with $N_{\mathrm{vib}}$ vibrational
states per electronic level, i.e.\ $N=2N_{\mathrm{vib}}$. The density matrix
$\rho$ is thus a $2N_{\mathrm{vib}}\times2N_{\mathrm{vib}}$ matrix with
equation of motion
\begin{equation}
   i \frac{d\rho}{dt}
   =
   \frac{1}{\hbar} 
   [H_{\mathrm{mol}} + H', \rho] 
   + 
   \mathcal{L}_{tr} \rho 
   +
   \mathcal{L}_{ph} \rho .
\label{eq:motion}
\end{equation}
The first term to the right governs the motion as a result of
the molecule Hamiltonian $H_{\mathrm{mol}}$, and the interaction $H'$,
between the molecule and the laser field.  The term
$\mathcal{L}_{tr}\rho$ yields the damping of the density matrix as a
result of transitions caused by $H_{\mathrm{fluct}}$ in which photons
are spontaneously emitted, but also transitions between different
vibrational levels are included in $\mathcal{L}_{tr}$ in a
phenomenological way.
The last term $\mathcal{L}_{ph}\rho$ makes it possible to introduce
additional phase relaxation, also in a phenomenological way. 
% Maybe the above is too much "talk" here.

The molecular Hamiltonian is diagonal, and can be written as a sum of
electronic and vibrational energies,
\begin{equation}
   H_{mol}
   =
   \sum_{n=0}^{N_{\mathrm{vib}}-1}
   \sum_{l=g,e} 
      |l;n>(E_l+ n\hbar\Omega_{\mathrm{vib}}) <l;n|.
\end{equation}
To see more clearly what $H'$ in Eq.\ (\ref{eq:motion}) means, let us
focus on a manageable example with just 4 levels
($N_{\mathrm{vib}}=2$). We then have
% \begin{equation}
%    \frac{H_{\mathrm{mol}}}{\hbar}
%    = 
%    \left[
%       \begin{array}{cccc}
%       0      &          0     &     0     &    0    \\
%       0      &  \omega_{\mathrm{vib}}  &  0     &    0    \\
%       0      &          0     &     \Omega_{ge}     &    0    \\
%       0      &          0     &     0     &
%       \Omega_{ge}+\omega_{\mathrm{vib}} \\
%       \end{array}
%    \right],
% \end{equation}
\begin{equation}
   \frac{H'}{\hbar}
   =
   \left[
      \begin{array}{cccc}
      0 & 0 & V_{-} f(0,0) & V_{-} f(0,1) \\
      0 & 0 & V_{-} f(1,0) & V_{-} f(1,1) \\
      V_{+} f(0,0) & V_{+} f(1,0) & 0 & 0 \\
      V_{+} f(0,1) & V_{+} f(1,1) & 0 & 0 \\
      \end{array}
   \right],
\end{equation}
and $H'$ is explicitly time-dependent since
\begin{equation}
   V_{\mp} = -p_0 E_0 e^{\pm i\Omega_L t}/2.
\end{equation}
With these equations, and their generalizations to cases with more
vibrational levels, the first term in Eq.\ (\ref{eq:motion}) can be
calculated.

Consider now the second term $\mathcal{L}_{\mathrm{tr}}\rho$, in Eq.\
(\ref{eq:motion}). Given a spontaneous transition rate $\Gamma_{kj}$
(at zero temperature) from level $j$ to level $k$, (in our model this
rate is due to vibrational damping for intraband transitions and due to
radiative damping and electron-hole pair creation for interband
transitions), this term can be written\cite{Scully}
\begin{widetext}
\begin{equation}
   \mathcal{L}_{tr}\rho
   =
   - 
   \sum_{jk}
   \langle n_{kj}+1 \rangle
   \frac{i\Gamma_{kj}}{2} 
   \left[
      \sigma_{jk}\sigma_{kj}\rho + \rho \sigma_{jk} \sigma_{kj} 
      - 2 \sigma_{kj} \rho \sigma_{jk}
   \right]
   - 
   \sum_{jk}
   \langle n_{kj} \rangle
   \frac{i\Gamma_{kj}}{2} 
   \left[
      \sigma_{kj}\sigma_{jk}\rho + \rho \sigma_{kj} \sigma_{jk} 
      - 2 \sigma_{jk} \rho \sigma_{kj}
   \right].
\label{eq:Ldamp}
\end{equation}
\end{widetext}
Here $\sigma_{ab}$ stands for an {\em operator} (or matrix) with the
$ab$ element equal to 1, and all the other elements equal to 0. 
% As the notation suggests, these operators are somewhat related to Pauli
% matrices. 
These operators fulfill the relations
\begin{equation}
   \sigma_{ab}\sigma_{kl} 
   =
   \sigma_{al} \delta_{bk},
\ \ \ \mathrm{and} \ \  \ 
   [\sigma_{ab}, \sigma_{kl}] 
   =
   \sigma_{al} \delta_{bk}
   -
   \sigma_{kb} \delta_{al}.
\label{eq:sigmamult}
\end{equation}
The first sum in Eq.\ (\ref{eq:Ldamp}) refers to (possibly thermally
activated) decay, the second to thermal excitation, and
\begin{equation}
   \langle n_{kj} \rangle = 
    1  / [e^{(E_j-E_k)/(k_BT)}-1],
\end{equation}
implying that the thermal bath consists of a number of harmonic
oscillators, something that is at least certainly true for the radiation
damping.  Given that all excitation energies $E_j-E_k$ occurring in
our model are fairly large compared with $k_BT$, we have restricted the
calculations to the zero-temperature limit, $\langle n_{kj} \rangle=0$.

% The transition rates used here are due to spontaneous photon emission
% from the excited electronic state to the ground state. By using the
% Fermi golden rule in the usual way, with $H_{\mathrm{fluct}}$ being the
% perturbation,  that rate is given  by
% \begin{equation}
%    \Gamma_{gn,em} = 
%    \frac{\omega^3}{3\pi \hbar \varepsilon_0 c^3} |p_0|^2 |f(n,m)|^2,
% \end{equation}
% where $\omega$ is the angular frequency corresponding to the transition
% energy, i.e.\
% \begin{equation}
%    \omega = 
%    \Omega_{ge} 
%    + (m-n) \omega_{\mathrm{vib}}.
% \end{equation}
% For vibrational relaxation, we use the parameter 
% $\gamma_{\mathrm{vib}}$ to represent the transition rate for transitions
% in which the electronic state of the molecule does not change while the
% vibrational quantum number decreases by 1 
% ($|n+1\rangle \rightarrow |n\rangle$).

The last term in the equation of motion describes dephasing of interband
(i.e.\ ground-excited and excited-ground) coherences.  This only
involves one density-matrix element at a time, thus
\begin{equation}
   \mathcal{L}_{ph}\,  \rho
   =
   \mathcal{L}_{ph}  \sum_{ij} \sigma_{ij}\, \rho_{ij} 
    =
    -i\gamma_{ph} 
    \sum_{i\in g}^{j\in e}
    (\sigma_{ij}\, \rho_{ij} + \sigma_{ji} \, \rho_{ji}),
\end{equation}
where $i\in g$ indicates that when the molecule is in state $i$ it
should be in the electronic ground state, etc.

\subsection{Stationary state} 

We need to solve Eq.\ (\ref{eq:motion}), and begin by
determining the stationary-state populations and coherences. 
% To this end we set the time-derivative on
% the left hand side of Eq.\ (\ref{eq:motion}) to zero and solve the
% resulting equation with the additional condition that $\Tr[\rho]=1$.
To facilitate the solution in a case with an arbitrary number of levels
we form an $N^2$-dimensional vector $\vec{\rho}$, from the elements of the
density matrix as
\begin{equation}
   \vec{\rho} =
   [\rho_{11}, \rho_{21}, \ldots \rho_{N1}, \rho_{12}, \ldots \rho_{NN}].
\end{equation}
The equation of motion can then be written 
\begin{equation}
   i\frac{d\vec{\rho}}{dt} 
   =
   \tensor{\mathcal{L}} \vec{\rho},
\end{equation}
where the tensor $\tensor{\mathcal{L}}$ describes the coupling between
two density matrix elements caused by the Hamiltonian and the damping.
$\tensor{\mathcal{L}}$ can be deduced from the right hand side of Eq.\
(\ref{eq:motion}). 

The stationary-state density matrix $\rho_{SS}$ is essentially
time-independent, but because of the explicitly time-dependent terms in
the Hamiltonian, it does have coherences (dipole moments)
oscillating with the laser frequency.   We therefore make the ansatz
\begin{equation}
   \vec{\rho}_{SS} =
   e^{-i\tensor{\Omega} t} 
   \vec{\rho}_{0},
\end{equation}
where $\vec{\rho}_0$ and $\tensor{\Omega}$ are time-independent, and
$\tensor{\Omega}$ is a diagonal tensor. Moreover,
the diagonal elements of $\tensor{\Omega}$ referring to
populations or intraband coherences equal 0. Only the diagonal
elements of $\tensor{\Omega}$  referring to interband coherences are
non-zero. They equal 
$+\Omega_L$, for ``excited-ground'' coherences and 
$-\Omega_L$ for ``ground-excited'' coherences.
For a two-state system (just electronic degrees of freedom) we would have
\begin{equation}
  \tensor{\Omega}
  =
   \left[
      \begin{array}{cccc}
      0 & 0 & 0 & 0 \\
      0 & \Omega_L & 0 & 0 \\
      0 & 0 & -\Omega_L & 0 \\
      0 & 0 & 0 & 0 
      \end{array}
   \right].
\end{equation}

The stationary state can thus be determined by solving the equation
\begin{equation}
   i\frac{d\rho_{SS}}{dt}
   =
   \tensor{\mathcal{L}} 
   \vec{\rho}_{SS}
   \ \ \Leftrightarrow \ \ 
   e^{-i\tensor{\Omega} t} 
   \tensor{\Omega}
   \vec{\rho}_{0}
   =
   \tensor{\mathcal{L}} 
   e^{-i\tensor{\Omega} t} 
   \vec{\rho}_{0},
\end{equation}
which after a multiplication by $e^{i\tensor{\Omega}t}$ can be rewritten as 
\begin{equation}
   [\tensor{\mathcal{L}}'-\tensor{\Omega} ]
   \vec{\rho}_{0} 
   =
   0,
\label{eq:matrix_stat}
\end{equation}
where 
\begin{equation}
   \tensor{\mathcal{L}}' = 
   e^{i\tensor{\Omega} t} 
   \tensor{\mathcal{L}}
   e^{-i\tensor{\Omega} t} 
\end{equation}
is time-independent. (Basically, we have transformed the equation of
motion to an interaction representation.) In addition to Eq.\
(\ref{eq:matrix_stat}), $\rho_0$ must satisfy
$\mathrm{Tr}[\rho_0]=1$.

\subsection{Calculation of the spectrum}

We can now focus on calculating the key quantities: the fluorescence
and Rayleigh and Raman scattering cross-sections, as given by  Eq.\
(\ref{eq:spectrum3}).  The positive frequency part of the dipole
operator originates from transitions from an electronically excited
state to the electronic ground state.  The negative frequency part, on
the other hand, comes from transitions from the electronic ground state
to the excited state.  With the aid of the $\sigma$ operators introduced
before Eq.\ (\ref{eq:sigmamult}) we can then write the expectation value
appearing in Eq.\ (\ref{eq:spectrum3}),
\begin{equation}
   \langle
      p^{(-)} (0)
      p^{(+)} (\tau)
   \rangle
   =
   |p_0|^2
   \sum_{b,k\in g}^{a,j\in e}
   f(b,a)  f(k,j)
   \langle
      \sigma_{ab}(0)
      \sigma_{kj}(\tau)
   \rangle .
\end{equation}
The expectation value 
$\langle \sigma_{ab}(0)  \sigma_{kj}(\tau) \rangle $
can be evaluated using the quantum regression theorem (QRT), as we will show
next.\cite{Scully} 

With the QRT
the value of any element of the density matrix at time $\tau$ can be
expressed as a linear combination of the density matrix elements at the
earlier time $0$, provided that the process is Markovian.
For a physical process to be Markovian there must not be any
memory effects, damping should be frequency-independent, and the
perturbing noise completely ``white.''  It is fairly clear that none of
these conditions are fulfilled in a strict sense here, for resonant
enhancement of the coupling between the molecule and the electromagnetic
field implies that damping and noise is stronger at some frequencies
than at others. However, the EM enhancement, see Fig.\ \ref{fig:Mfreq},
varies relatively slowly, on an energy scale $\Delta E$ of a few
tenths of eV, corresponding to a memory time scale $\hbar/\Delta E$ of a
few femtoseconds, shorter than the other time scales relevant for the
molecule dynamics. At the same time we note that in a number of
interesting problems in quantum optics, for example an atom in a
photonic crystal with an atomic transition energy near a band-gap edge
of the photonic crystal, involves very important memory effects
that require other theoretical methods and leads to  novel physical
phenomena.\cite{John:94,JackHope:01}

Continuing the calculation,
we first consider a one-operator expectation value and  note that 
\begin{equation}
   \langle \sigma_{kj}(\tau) \rangle
   =
   \Tr[\rho(\tau) \sigma_{kj}] = \rho_{jk}(\tau).
\end{equation}
We express this dependence through a Green's
function 
$G_{jk,rs}(\tau)$, thus
\begin{equation}
   \rho_{jk}(\tau) 
   =
   \sum_{rs} G_{jk,rs}(\tau) \rho_{rs}(0).
\label{eq:rhogreen}
\end{equation}
In the previously used vector and tensor notation this equation would read 
$
   \vec{\rho}(\tau)
   =
   \tensor{G}(\tau) \vec{\rho}(0)
$.
In terms of operator expectation values Eq.\ (\ref{eq:rhogreen}) means
\begin{equation}
   \langle \sigma_{kj}(\tau) \rangle
   =
   \sum_{rs} G_{jk,rs}(\tau) 
   \langle \sigma_{sr}(0) \rangle.
\label{eq:expectgreen1}
\end{equation}
%  Are we going in circles here?
The quantum regression theorem\cite{Scully} 
states that the relation between the
two-operator correlation functions 
$\langle \sigma_{ab}(0) \sigma_{kj}(\tau) \rangle$
and 
$\langle \sigma_{ab}(0) \sigma_{sr}(0) \rangle$
is identical to the one between one-operator expectation values
expressed by Eq.\ (\ref{eq:expectgreen1}), thus
\begin{equation}
   \langle \sigma_{ab}(0) \sigma_{kj}(\tau) \rangle
   =
   \sum_{rs} G_{jk,rs}(\tau) 
   \langle \sigma_{ab}(0) \sigma_{sr}(0) \rangle.
\label{eq:expectgreen2}
\end{equation}
It will now be possible to calculate the dipole-moment correlation
function since the expectation value appearing in Eq.\
(\ref{eq:expectgreen2}) can be evaluated from the stationary-state
density matrix $\rho_{SS}$, and the Green's function, in view of Eq.\
(\ref{eq:rhogreen}), can be deduced from the
equation of motion for the density matrix.

To explicitly calculate the Green's functions  we need to solve the
equations of motion for the density matrix given certain initial
conditions at time $t=0$, since 
obviously Eq.\ (\ref{eq:rhogreen}) states that $G_{ji,kl}(\tau)$ is
the value that the element $\rho_{ji}$ will take at time $\tau$ given
that all density matrix elements are 0 at time $t=0$ except the
$\rho_{kl}$ which equals 1. 
[This should be viewed strictly
mathematically; the initial conditions here sometimes mean that the
density matrix is traceless and non-Hermitian. However, when
$\tensor{G}$ operates on a physical $\vec{\rho}(0)$
one obtains a sensible result also for $\vec{\rho}(\tau)$.]
We can thus write down an
equation of motion for the Green's tensor
\begin{equation}
   i \frac{d\tensor{G}}{dt}
   -
   \tensor{\mathcal{L}} \tensor{G}
   =
   i \tensor{1} \delta(t),
\label{eq:greenmotion}
\end{equation}
where the right hand side takes care of the initial conditions [i.e.\
$\tensor{G}(t)\equiv0$ for $t<0$]. 

To solve Eq.\ (\ref{eq:greenmotion}) we make an ansatz analogous
to the one made for $\rho_{SS}$ above,
\begin{equation}
   \tensor{G}(t) 
   =
   e^{-i\tensor{\Omega}t}
   \tensor{G}_0(t).
\label{eq:greenansatz}
\end{equation}
% By inserting this into Eq.\ (\ref{eq:greenmotion}) and multiplying by 
% $e^{i\tensor{\Omega}t}$ from the left we get
% \begin{equation}
%  i
%  \frac{d\tensor{G}_0}{dt}
%  +
%  \tensor{\Omega} \tensor{G}_0 
%  -
%  \tensor{\mathcal{L}}' \tensor{G}_0 
%  =
%  i \tensor{1} \delta(t).
% \end{equation}
Then by introducing the Fourier transform $\tensor{G}_0(\omega)$
of $\tensor{G}_0(t)$ through
\begin{equation}
   \tensor{G}_0(\omega)
   =
   \int_{-\infty}^{\infty} dt\,
   e^{i\omega t} \, 
   \tensor{G}_0(t),
\end{equation}
% $\tensor{G}_0(\omega)$ can be shown to follow the equation 
% \begin{equation}
%  \omega \tensor{1} \tensor{G}_0(\omega)
%  +
%  \tensor{\Omega} \tensor{G}_0(\omega)
%  -
%  \tensor{\mathcal{L}}' \tensor{G}_0(\omega)
%  =
%  i \tensor{1},
% \end{equation}
we arrive at the solution 
\begin{equation}
   \tensor{G}_0(\omega) 
   =
   i
   \left[
      (\omega+i\delta) \tensor{1}
      +
      \tensor{\Omega}
      -
      \tensor{\mathcal{L}}'
   \right]^{-1}.
\end{equation}
From a formal point of view the imaginary part $\delta=0$, but to avoid
that the solution diverges at the driving frequency $\Omega_L$ we
introduce a finite, but small,  $\delta$.  This gives a width to the
calculated Rayleigh scattering peak. In the calculations reported in
this paper we have used the value $\delta=2\times10^{12} \unit{s^{-1}}$
unless otherwise stated.

We can now insert the expression for the expectation value into the
equation for the cross-section,
\begin{eqnarray}
   &&
   \frac{d^2\sigma}{d\Omega d(\hbar\omega)}
   =
   \frac{\omega^4 |M(\omega, \theta)|^2}
   {I_{in} 8 \pi^3 c^3 \varepsilon_0\hbar}
   |p_0|^2
   \sum_{b,k\in g}^{a,j \in e}
   f(b,a)\,f(k,j)\,
   \breakeq
   \times
   \Real
   \int_{0}^{\infty} d\tau\, e^{i\omega\tau}
   \sum_{rs}
   G_{jk,rs}(\tau)
   \langle \sigma_{ab}(0) \sigma_{sr}(0) \rangle.
\end{eqnarray}
The time integral gives the Green's function Fourier transform.
In view of Eq.\ (\ref{eq:greenansatz}) and the fact that the
index $j$ refers to an excited state while $k$ refers to the electronic
ground state the time integral can be written
\begin{equation}
   \int_{0}^{\infty} d\tau\, e^{i\omega\tau}
   e^{-i\Omega_L \tau} G_{0jk,rs}(\tau)
   =
   G_{0jk,rs}(\omega-\Omega_L).
\end{equation}
Furthermore, the expectation value can be simplified to
\begin{equation}
   \langle \sigma_{ab}(0) \sigma_{sr}(0) \rangle
   = \langle \sigma_{ar} (0) \rangle  \delta_{bs}
   = \rho_{0,ra} \delta_{bs},
\end{equation}
which finally yields
\begin{eqnarray}
   &&
   \frac{d^2\sigma}{d\Omega d(\hbar\omega)}
   =
   \frac{\omega^4 |M(\omega, \theta)|^2}
   {I_{in} 8 \pi^3 c^3 \varepsilon_0\hbar}
   |p_0|^2
   \sum_{b,k\in g}^{a,j \in e}
   f(b,a)\,f(k,j)\,
   \breakeq
   \times
   \sum_{r}
   \rho_{0,ra}
   \Real
   [
      G_{0jk,rb}(\omega-\Omega_L)
   ]
\label{eq:final}
\end{eqnarray}
for the scattering and fluorescence cross-section.

\subsection{Results in an elementary case}

Equation (\ref{eq:final}) contains contributions to the cross section
due to both diagonal and off-diagonal elements of the density matrix.
These contributions typically have very different physical origins. The
light emission related to diagonal elements are due to an excited state
being populated before decaying; this describes fluorescence
processes. The off-diagonal elements, on the other hand, represent an
oscillating dipole moment on the molecule, and the corresponding
contributions to the emitted light are due to various scattering
processes.

\begin{figure}[tb]
   \includegraphics[angle=0,width=8.0 cm]{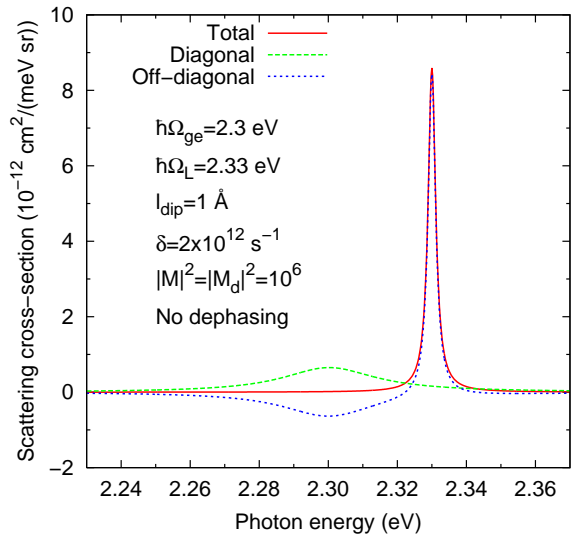}
   \caption{(color online)
      The spectrum of scattered light from a two-level system. In
      addition to showing the total cross section with a Raleigh peak as
      calculated from Eq.\ (\ref{eq:final}), the other two curves show
      the contributions originating from the diagonal and off-diagonal
      elements of the density matrix, respectively. 
   }
   \label{fig:diag}
\end{figure}

To illustrate this we show results for the scattering cross section in a
very simple case in Fig.\ \ref{fig:diag}. We consider a model molecule
with only two electronic levels without any sublevel structure due to
vibrations (cf.\ Ref.\ \onlinecite{Mollow}).  The electronic excitation
energy is set to 2.3 eV and the incident laser photon energy is 2.33 eV.
The electromagnetic enhancement is used as a parameter: both $|M|^2$
and $|M_d|^2$ are set to $10^6$, independent of the frequency $\omega$.
As for damping mechanisms, we of course keep the radiation damping, but
there is no dephasing in this model. By inspecting Eq.\ (\ref{eq:final})
we see that indices $b$, $k$, $a$, and $j$ in this case refer to one
definite state (whereas they in the general case run over a set of
different vibrational states), only the index $r$ can point to either
the ground state or the excited state, so the cross section is a sum of
two terms. The three curves in Fig.\ \ref{fig:diag} show the total cross
section as well as the contributions from the term proportional to the
element $\rho_{0,ee}$ (diagonal) and the term proportional to
$\rho_{0,ge}$ (off-diagonal).  The total cross-section is everywhere
positive as it should, and has a sharp peak at $\hbar\omega=2.33
\unit{eV}$ due to Rayleigh scattering, (the width of this peak is set by
the parameter $\delta$ used in the calculation).  The Rayleigh
scattering is of course the result of photons being re-emitted because
the molecule has an oscillating dipole moment described by the
off-diagonal elements of the density matrix, and this curve follows the
one for the full cross-section closely near the laser frequency.
However, for photon energies near the electron transition energy the
off-diagonal contribution becomes negative, largely canceling the
diagonal contributions in this case.

\section{Calculated spectra}
\label{sec:results}

\begin{figure}[tb]
   \includegraphics[angle=0,width=8.0 cm]{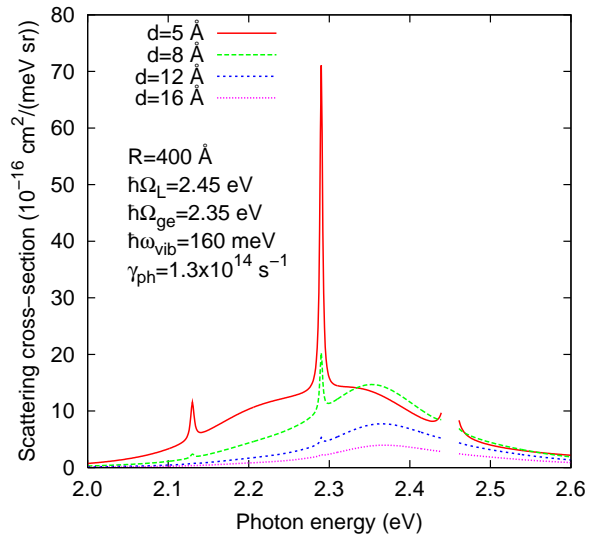}
   \caption{(color online)
      Calculated fluorescence and Raman spectra for a molecule placed
      symmetrically $d=d_1=d_2$ between two spherical ($R=400$ \r{A}) Ag
      nanoparticles. Different curves correspond to different
      particle-molecule separations $d$. 
      The molecule parameter values
      used are $\hbar\Omega_L=2.45$ eV, $\hbar\omega_{\mathrm{vib}}=160
      \unit{meV}$, $\gamma_{ph}=1.3\times10^{14} \unit{s^{-1}}$,
      $\gamma_{\mathrm{vib}^{(g)}}=2\times10^{12} \unit{s^{-1}}$,
      $\gamma_{\mathrm{vib}^{(e)}}=10\times10^{12} \unit{s^{-1}}$,
      $\alpha=0.5$,
      and $I_{in} \approx 0.13 \unit{\mu W/(\mu m^2)}$.
   }
   \label{fig:dspectra}
\end{figure}

We begin by looking at spectra calculated for a molecule placed
symmetrically between {\em two} spherical Ag nanoparticles.
Figure \ref{fig:dspectra} displays a number of combined
fluorescence and Raman spectra corresponding to different
molecule-particle separations $d$, and thereby different electromagnetic
enhancements. 
The plotted differential cross section corresponds to a situation where
both the incident and scattered light propagate in the symmetry plane
(i.e.\ $\theta=90\mgrader$).
All the spectra show a broad fluorescence background.
For the smaller values of $d$, sharp Raman peaks rise above this
background. The fundamental Stokes  peaks, red-shifted by
$\hbar\omega_{\mathrm{vib}}$ from the laser photon energy
$\hbar\Omega_L$, are the highest ones, but at least for $d=5$ \r{A} we
can also see an overtone peak at
$\hbar\Omega_L-2\hbar\omega_{\mathrm{vib}}$ due to creation of multiple
vibrational excitations. The very strong dependence of the Raman peaks
on the molecule-particle separation $d$ is the most striking tendency
seen in the plot.  Increasing $d$ from 5 \r{A} to 8 \r{A} reduces the
peak height by approximately a factor of 6, at $d=12$ \r{A} only a small
Raman peak remains, and at $d=16$ \r{A} it is nearly impossible to
discern a Raman peak. The main reason for this is the stronger EM
enhancement one gets with a smaller $d$, cf.\ Figs.\ \ref{fig:Mfreq},
and \ref{fig:Md}.  As was shown in Fig.\ 3 of Ref.\ \onlinecite{Xu:04},
the Raman scattering cross section behaves as $\sigma_R \approx
|M(\Omega_L)|^2|M(\Omega-\omega_{\mathrm{vib}})|^2
\sigma_{R,\mathrm{free}}$ as long as $d$ is not too small. The Raman
cross section scales with the fourth power of the enhancement because
Raman scattering involves two steps: in the first step a photon is
temporarily absorbed and the molecule goes into a virtual state, in the
second step a photon is emitted while the molecule goes back to the
ground state, albeit to a different vibrational state.  The rate of both
these steps are enhanced by a factor $|M|^2$. 

Also the fluorescence background in Fig.\ \ref{fig:dspectra} changes
with $d$; the fluorescence cross section shows an increasing tendency
with decreasing $d$, however, this change is not at all as marked as for
the Raman signal.  Naively one may think that fluorescence, which
involves an absorption event and an emission event, should also display
a cross section scaling as $|M|^4$.  But this is not so, because in
the case of fluorescence the molecule is in a real, electronically
excited state after having absorbed a photon but before emitting the
fluorescence photon.  Looking at the final step of the fluorescence
process it is then clear that there are two factors that determine the
fluorescence intensity: (i) the EM enhancement for the frequency of the
emitted photon $|M(\omega)|^2$, and (ii) the probability of finding
the molecule in the excited electronic state $P_{\mathrm{exc}}$.  The
probability $P_{\mathrm{exc}}$ is affected by the electromagnetic
enhancement, but in two competing ways that largely tend to cancel each
other.  $P_{\mathrm{exc}}$ increases when the molecule is excited from
the electronic ground state and the rate of such processes scales as
$|M(\Omega_L)|^2$. However, at the same time the excited state is
emptied by radiative (including fluorescence) and
non-radiative processes, and the rate of these is given by
$|M_d|^2 \gamma_{0,\mathrm{rad}}P_{\mathrm{exc}}$.  Consequently, the
probability of finding the molecule in an electronically excited state
depends in a stationary-state situation on the EM enhancement factors as 
$$
   P_{\mathrm{exc}}
   \sim
   |M|^2/|M_d|^2, 
$$
and for the fluorescence cross section we get 
$$
   \sigma_F 
   \sim
   |M|^2 P_{\mathrm{exc}} 
   \sim
   |M|^4 / |M_d|^2.
$$

\begin{figure}[tb]
   \includegraphics[angle=0,width=8.0 cm]{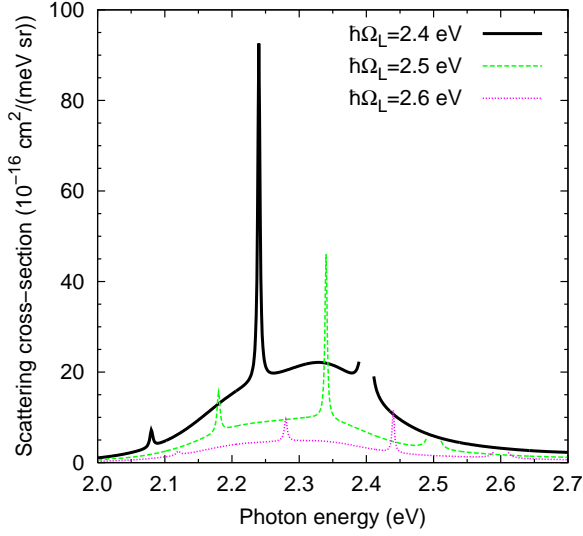}
   \caption{(color online)
      Calculated spectra for three different frequencies of the driving
      field. The molecule-particle separation is $d=5$ \r{A}, and the
      rest of the parameter values are the same as in Fig.\
      \ref{fig:dspectra}.
   }
   \label{fig:Ospectra}
\end{figure}

In Fig.\ \ref{fig:Ospectra} we compare three spectra calculated with
different laser photon energies. These spectra result from the combined
effect of a frequency-dependent Raman cross section, see Fig.\
\ref{fig:profiles} (b), and the frequency-dependence of the
electromagnetic enhancement as shown in Fig.\ \ref{fig:Mfreq}. The
free-molecule Raman cross section is considerably higher for both
$\hbar\Omega_L=2.4 \unit{eV}$ and $\hbar\Omega_L=2.5 \unit{eV}$ than for
$\hbar\Omega_L=2.6 \unit{eV}$, which explains why the Raman peak in the
latter case is so small. In addition, as a result of the resonant maximum
at $\approx 2.2$ eV, in Fig.\ \ref{fig:Mfreq}, the combined EM
enhancement $|M(\Omega_L-\omega_{\mathrm{vib}})|^2 |M(\Omega_L)|^2$ is
the largest for $\hbar\Omega_L=2.4$ eV and smallest for
$\hbar\Omega_L=2.6$ eV.

\begin{figure}[tb]
   \includegraphics[angle=0,width=8.0 cm]{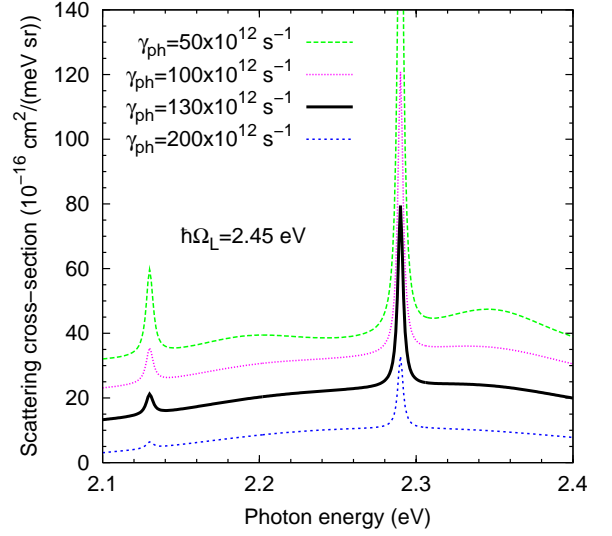}
   \caption{(color online)
      Calculated spectra for four different dephasing rates.
      The molecule-particle separation is $d=5$ \r{A}, and the
      rest of the parameter values are the same as in Fig.\
      \ref{fig:dspectra}. The curves have been shifted vertically for
      increased clarity.
      The truncated peak for $\gamma_{ph}=50\times10^{12} \unit{s^{-1}}$
      has height $\approx 200\times10^{-16} \unit{cm^2/(meV\,sr)}$.
   }
   \label{fig:gspectra}
\end{figure}
% 'HOME/Progs/raman03/jrsres2/series5/gamma50/ramanspect.dat' \

Figure \ref{fig:gspectra} shows the spectra that result when the
dephasing rate is varied. The diagram only shows the portion of the
spectrum where the Raman peaks (the principal one and the first
overtone) falls. Both these cross sections are, at least with a laser
photon energy as close to resonance as 2.45 eV, quite sensitive to the
dephasing rate.   This is particularly apparent for the Raman overtone
peak. 
For creation of multiple vibrational excitations to take place, the
molecule must have a chance to ``investigate'' the displaced oscillator
potential during a longer time than in the case of
single-excitation creation.  This means that the multiphonon processes
are more easily disrupted when the dephasing rate goes up.

\begin{figure}[tb]
   \includegraphics[angle=0,width=8.0 cm]{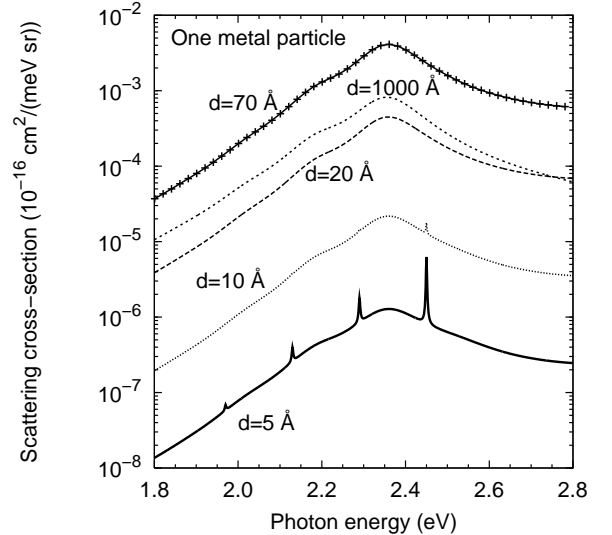}
   \caption{
      Calculated spectra when the molecule is placed near only one Ag
      particle with radius 400 \r{A}.  The molecule parameter values are
      the same as in Fig.\ \ref{fig:dspectra}.
   }
   \label{fig:onespectra}
\end{figure}

In Fig.\ \ref{fig:onespectra} we show results for a different
nanoparticle configuration than the one considered so far.  The results
here refer to a molecule placed near only one silver particle.  In this
case there is some electromagnetic enhancement $M$, because the metal
particle can be (resonantly) polarized. But due to the lack of
electromagnetic particle-particle coupling the enhancement has a much
smaller magnitude than in the two-particle case. Typically $|M|^2$
reaches values of 20--40 (depending on frequency) for $d=5$ \r{A}.  $M$
is less sensitive to $d$ than in the two-sphere case, the most important
contribution coming from the dipole field around the particle, i.e.\
$|M|\sim R^3/(R+d)^3$.  However, the damping rates due to losses in the
particles, especially electron-hole pair creation, are still comparable
to those in the two-particle case.  This means that fluorescence is very
strongly suppressed for small $d$ when there is only one Ag particle.
For example, for $d=5$ \r{A} a rough estimate of the ratio
$|M|^4/|M_d|^2$ with $|M|^2\approx 30$ and $|M_d|^2$ taken from Fig.\
\ref{fig:Md} shows that we can expect a suppression (or quenching) of
the fluorescence by a factor of $\sim 10^{3}$ compared with the
free-molecule case.  In Fig. \ref{fig:onespectra} we see in fact a
difference by about 3 orders of magnitude between the fluorescence when
$d$ is 5 \r{A} and when $d$ is 1000 \r{A}, the latter $d$ yields
situation quite similar to  the free-space case.  For $d=5$ \r{A} the
fluorescence is suppressed to the extent that Raman peaks stand out from
the background, however, the absolute Raman cross sections are of course
too small to be detectable in an experiment with a single molecule.

With increasing $d$ in Fig.\ \ref{fig:onespectra}, the fluorescence
yield first increases because $|M_d|$ decreases rapidly whereas $|M|$
falls off at a much slower rate.  Then when $d$ reaches values of
50--100 \r{A} (see the 70-\r{A} curve) the fluorescence has a maximum,
because the rate of decrease in $|M|^4$ overtakes that of
$|M_d|^2$.  For even larger $d$ $|M|^2\approx|M_d|^2$, so the
fluorescence cross section behaves as $|M|^2$ there. The spectrum
calculated for $d=1000$ \r{A} approaches what one finds for the molecule
in free space.

\begin{figure}[tb]
   \includegraphics[angle=0,width=8.0 cm]{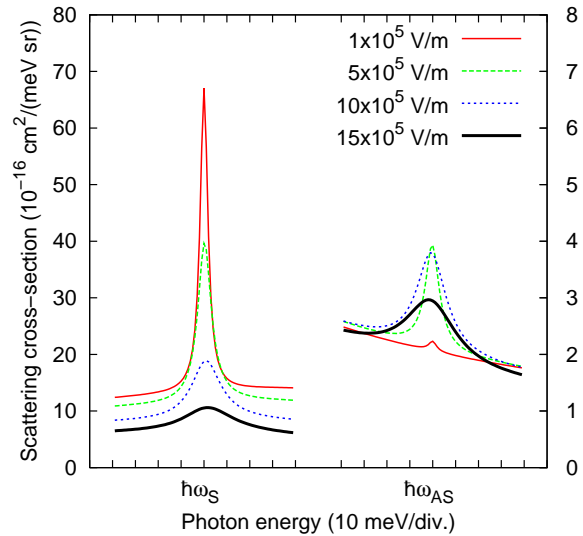}
   \caption{(color online)
      Calculated spectra for a molecule placed between two nanoparticles,
      with $d=5$ \r{A}, for a number of different incident intensities. The
      other parameter values are the same as in Fig.\
      \ref{fig:dspectra}.
      Only the energy ranges around the Raman Stokes peak at
      $\hbar\omega_{S}=2.29 \unit{eV}$ and the Raman anti-Stokes peak
      at $\hbar\omega_{AS}=2.61 \unit{eV}$ are shown. 
   }
   \label{fig:highfield}
\end{figure}

The results presented so far have been calculated with moderate laser
intensities.  We have used the value $10^4 \unit{V/m}$ for $E_0$ in Eq.\
(\ref{eq:laserfield}) which corresponds to 
$I_{\mathrm{in}}  \approx 0.13 \unit{\mu W/(\mu m^2)}$. This intensity
is so low that, even in spite of the EM enhancement, the molecule spends
almost all the time in the ground state (assuming, as we have done
before, that thermal excitations can be neglected).
The results presented in Fig.\ \ref{fig:highfield} (where we return to a
situation with two nanoparticles)
have been calculated
with considerably higher incident intensities. This means that we encounter
situations where there is an appreciable probability of finding the molecule
electronically and/or vibrationally excited. For very strong incident
fields also the response properties of the molecule changes, essentially
because the probability of finding the system in a certain level becomes
time-dependent (as a result of Rabi oscillations).  In our calculations
the stationary-state density matrices $\rho_{SS}$ and $\rho_0$ describe
a time-average, the way the molecule ``looks'' on the average after long time.
The dipole-dipole correlation function 
$\langle p^{(-)}(0)p^{(+)}(t) \rangle$, 
on the other hand, contains information about the molecule
dynamics over a shorter period of time, starting from a certain initial
state.  Here Rabi oscillations due to excitation and deexcitation by a
strong driving, external field show up in the results, along with other,
more apparent, time-dependent aspects of the molecular dynamics such as dipole
oscillations and damping.

%  Fields and intensities....
% 1  13 muW/mum^2
% 2  53
% 3  119
% 4  212
% 5  332
% 6  478
% 7  650
% 8  849
% 9 1075
% 10 1328
% 11 1607
% 12 1912
% 13 2244
% 14  2603
% 15  2988

In Fig.\ \ref{fig:highfield}, we restrict the attention to parts of
the spectra in the  frequency ranges around the Stokes and
anti-Stokes peaks with center frequencies $\omega_S=
\Omega_{ge}-\omega_{\mathrm{vib}}$ and $\omega_{AS} =
\Omega_{ge}+\omega_{\mathrm{vib}}$, respectively.
The different spectra correspond to intensities ranging from $\approx 13
\unit{\mu W/ (\mu m^2)}$ (at $E_0=10^5 \unit{V/m}$) 
to $\approx 3.0 \unit{mW/(\mu m^2)}$ (at $E_0=15\times10^5 \unit{V/m}$).
The rest of the parameter values have been chosen the same way as in
the calculation behind Fig.\ \ref{fig:dspectra}. At the lowest
intensity we get a Stokes peak that is nearly identical to the one in
Fig.\ \ref{fig:dspectra}, but with increasing intensity the peak height
diminishes quite rapidly, and it is also broadened.  At the same time,
at the anti-Stokes frequency, a small bump for $E_0=10^5 \unit{V/m}$ 
develops into a marked peak at $E_0=5\times10^5 \unit{V/m}$, and
subsequently also this peak is reduced in height and broadened with
increasing $E_0$. Thus, with incident intensities of the order of 
$1 \unit{mW/\mu m^2}$ it is possible to obtain anti-Stokes Raman
scattering (in the model) with a cross section that is experimentally
detectable.  The peak values found in the figure must be multiplied by
an effective peak width $\sim 10 \unit{meV}$ and the effective solid angle 
(for
dipole scattering) $8\pi/3$ to give the total Raman anti Stokes cross
section.  The fact that the anti-Stokes signal is weaker than the Stokes
signal is here partly due to the choice of laser frequency which means
that the Stokes peak falls at a maximum in the EM enhancement while the
anti-Stokes peak is at a minimum for $|M|^2$.

\begin{figure}[tb]
   \includegraphics[angle=0,width=8.0 cm]{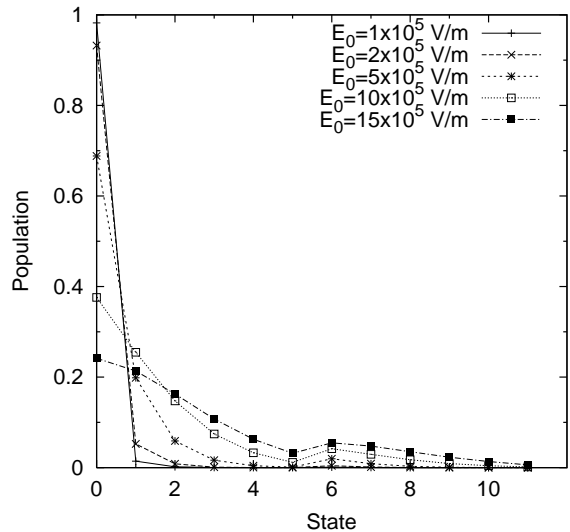}
   \caption{
      The probability of finding the molecule in different states and
      for a number of different driving fields. States numbered 0
      through 5  on the horizontal axis are in the electronic ground
      state, i.e.\ $| g \rangle 0; n \rangle$ for $0 \le n \le 5$,
      whereas states 6 through 11 correspond to the electronically
      excited states  
      $| e \rangle x_0; 0 \rangle$ 
      through
      $| e \rangle x_0; 5 \rangle$.
   }
   \label{fig:population}
\end{figure}

To discuss the tendencies in Fig.\ \ref{fig:highfield} as a function of
the incident field it is useful to calculate the probabilities of
finding the molecule in different states. This is shown in  
Fig.\ \ref{fig:population}.
The development at moderate intensities can be well understood just from
looking at the probabilities of finding the molecule in either of the two
lowest states 
$|g\rangle |0;0\rangle$
and 
$|g\rangle |0;1\rangle$. 
With increasing intensity it becomes more likely to find the molecule in
the vibrationally excited state which is a good initial state for the
anti-Stokes Raman process. At the same time the vibrational ground
state, the most common initial state for a Stokes Raman process, becomes
more and more depopulated leading to a decreased intensity there. 
However, comparing the probabilities in Fig.\ \ref{fig:population} with
the changes in peak heights in Fig.\ \ref{fig:highfield}, it becomes
clear that the behavior at higher intensities cannot be understood
exclusively in terms of the probabilities.  If this was true, the Stokes
peak would not diminish as rapidly as it does, and the anti-Stokes peak
would continue to increase in height. The reduced strength of both the
anti-Stokes and Stokes peaks is instead the result of the molecule
response becoming time-dependent at strong driving fields which leads
to a smearing-out of all spectral features.

% SUMMARY SECTION
\section{Discussion and Summary}
\label{sec:summary}

We have performed a theoretical analysis of surface-enhanced Raman
scattering and fluorescence based on a treatment that combines
electromagnetic enhancement effects and molecule dynamics. We used a
simple molecule model, however, with parameter values chosen in a
realistic way.  

The calculated results give Raman scattering cross sections of order
$10^{-14} \unit{cm^2}$, thus comparable to the values found in single
molecule SERS experiments,\cite{Nie:97}  provided the laser frequency is
reasonably close to an electronic excitation frequency of the molecule,
and the electromagnetic environment (the nanoparticles) provide an EM
enhancement of the Raman cross section by some 10 orders of magnitude.
Thus, we conclude that the 14 to 15 orders of magnitude enhancement
sometimes cited in connection with single-molecule SERS may neither be
present nor needed for the effect to occur.

Our study gives an opportunity to study how the Raman and fluorescence
parts of the spectra develop as the electromagnetic enhancement is
varied. Most of the trends found here agree well with what one can
expect compared with experiment. For the highest enhancements the Raman
peaks stand out from the fluorescence background. The rather rapid
decrease of the Raman cross section with increasing molecule-particle
separation agrees qualitatively with results found in experiments where
the metal particles are coated with a thin layer of dielectric
material, the thickness of which can be controlled with reasonable
precision.\cite{Bao:04,Cotton:98}
At the same
time the fluorescence background is more structured here than in most
experimental results found in the literature,\cite{Nie:97,Xu:99} 
although some exceptions do exist.\cite{Weiss:01} 

% Need some Drexhage reference here.
The quenching of the fluorescence (Fig.\ \ref{fig:onespectra}) once a
molecule placed near a metal surface is a well-known phenomenon, and
systematic studies have also found a local maximum in the
fluorescence cross section as a function of the molecule-surface
distance.\cite{Kreiter:04} Compared with this study, the fluorescence
maximum occurs closer to the surface in our results because of the
faster drop of the electromagnetic enhancement with increasing distance
from a finite nanoparticle than from a flat surface.

The model we have used in this work is a basic one containing a minimum
of ingredients.  To better account for some aspects of the SERS
phenomenon, further developments are needed: (i) More vibrational modes
should be added to better describe a fairly large organic molecule. (ii)
More electronic levels could also be included. This could be a way to
model ``blinking'' phenomena due to the molecule's  spending some time
in a metastable state that is not directly optically active.  One could
also introduce electronic states leading to dissociation of the
molecule, thus modeling photobleaching effects.  (iii) Ultimately one
should also try to include electron transfer processes between the
molecule and the metal particles.  

% Last point at least need polishing.

\begin{acknowledgments}
   We thank Hongqi Xu, Xue-Hua Wang, Martin Persson,
   and Lars Samuelson for stimulating discussions.  This work was
   supported by the Swedish Research Council (VR) and the Swedish
   Foundation for Strategic Research (SSF).  Part of this work was
   carried out while one of us (P.J.) visited the National Institute of
   Material Science in Tsukuba, Japan and IEAP at the Christian-Albrechts
   Universit\"at in Kiel, Germany. The support from Dr. Zhen-Chao Dong
   in Tsukuba and Prof.\ Richard Berndt and his group in Kiel is
   gratefully acknowledged.
\end{acknowledgments}

\appendix  % yields numbered (A, B, C,...) Appendixes.
% Use \appendix* to suppress numbering (just one Appendix)

\section{Electromagnetic calculation}
\label{app:em}

The coefficients $\tilde{c}$ in Eq.\ (\ref{eq:asystem}) are 
explicitly given by 
\begin{equation}
   \tilde{c}_{\tau\ell,m,\tau'\ell',m'}(\vec{R})
   =
   \sum_{j=-1}^{1}
   F_{\ell m}^{j} 
   F_{\ell' m'}^{j} 
   \tilde{A}_{\ell,m+j,\ell', m'+j}(\vec{R}),
\label{eq:tildec}
\end{equation}
when $\tau=\tau'$, and
\begin{eqnarray}
   &&
   \tilde{c}_{\tau\ell,m,\tau'\ell',m'}(\vec{R})
   =
   \frac{2\ell' + 1}{i\sqrt{\ell'(\ell'+1)}} \,
   \sum_{j=-1}^{1}
   \mathcal{R}_{-j}  \,
   \breakeq
   \times
   F_{\ell m}^{j}\, 
   \tilde{A}_{\ell,m+j,\ell'-1, m'+j}(\vec{R}) \,
   C_{\ell',m',\ell'-1,m'+j}^{1,-j},
\end{eqnarray}
when $\tau=\overline{\tau'}$.
In these expressions
\begin{equation}
   F_{\ell m}^{0} 
   =
   \frac{m}{\sqrt{\ell(\ell+1)}}
   \ \ 
   \mathrm{and}
   \ \ 
   F_{\ell m}^{\pm 1} 
   =
   \frac{\sqrt{(\ell \mp m)(\ell \pm m+1)}}{\sqrt{2\ell(\ell+1)}},
\end{equation}
\begin{equation}
   \mathcal{R}_{+1} = - \sqrt{4\pi/3},
   \ \ \mathrm{while} \ \ 
   \mathcal{R}_{0} = 
   \mathcal{R}_{-1} =  
   \sqrt{4\pi/3},
\label{eq:hatRcoeff}
\end{equation}
and the Gaunt integrals  $C_{\ell,m,\ell',m'}^{LM}$ are defined by
\begin{equation}
   C_{\ell,m,\ell',m'}^{LM}
   =
   \int d\Omega 
   Y_{\ell m}^{*} (\Omega)
   Y_{LM} (\Omega)
   Y_{\ell' m'} (\Omega).
\end{equation}
The coefficients $\tilde{A}$ in Eq.\ (\ref{eq:tildec}) are given by
\begin{eqnarray}
   \tilde{A}_{\ell,m,\ell',m'}(\vec{R})
   &&
   =
   4 \pi
   \sum_{L=|\ell-\ell'|}^{\ell+\ell'} 
   (-i)^{\ell - L -\ell'}\, 
   h_{L}(k|\vec{R}|) 
   \breakeq
   \times
   Y_{L,m-m'}(\hat{R})\, 
   C_{\ell,m,\ell',m'}^{L,m-m'}.
\end{eqnarray}
Finally, the coefficients describing the external field in Eq.\
(\ref{eq:planewave}) are 
\begin{equation}
   a_{1\ell m,\mathrm{ext}}^{s} 
   =
   4\pi i^{\ell} 
   e^{i \vec{k}\cdot\vec{r}_s}
   \sum_{j=-1}^{1}
   Y_{\ell,m+j}^{*}(\hat{k}) \, F_{\ell m}^{j}\,
   [\hat{e}_j^* \cdot \vec{E}_0],
\end{equation}
and 
\begin{equation}
   a_{2\ell m,\mathrm{ext}}^{s} =
   4\pi i^{\ell} 
   e^{i \vec{k}\cdot\vec{r}_s}
   \sum_{j=-1}^{1}
   Y_{\ell,m+j}^{*}(\hat{k}) \, F_{\ell m}^{j}\,
   \left[\hat{e}_j^* \cdot 
      (\hat{k}\crossprod \vec{E}_0)\right],
\end{equation}
where the unit vectors 
$\hat{e}_{-1}$,
$\hat{e}_{0}$,
and
$\hat{e}_{1}$
are given by
\begin{equation}
   \hat{e}_0 = \hat{z} 
   \ \ 
   \mathrm{and}
   \ \ 
   \hat{e}_{\pm 1} = (\hat{x} \pm \hat{y}/i)/\sqrt{2},
\end{equation}
and $\hat{k}=\vec{k}/|\vec{k}|$ is a unit vector in the direction of
$\vec{k}$.
When instead the spheres are driven by a dipole 
$\hat{z}p_0e^{-i\omega t}$ at the molecule position we have
\begin{equation}
   a_{\tau\ell m,\mathrm{ext}}^{s} =
   \frac{p_0k^3}{4\pi\varepsilon_0}\,
   \sqrt{\frac{8\pi}{3}}\,
   \tilde{c}_{2,1,0,\tau,\ell,m} (\vec{r}_s).
\end{equation}

\section{Electron-hole pair damping}
\label{app:eh}

In this Appendix we outline the calculation of the electron-hole pair
contribution $P_{eh}/P_{\mathrm{free}}$ to the damping rate enhancement
$|M_d|^2$.  The non-local dielectric response of the metal particles is
treated within d-parameter theory,\cite{Liebsch:87} and the derivation
to a large extent follows Ref.\ \onlinecite{Image}.

The idea is to model the interaction between the molecular dipole and
the degrees of freedom (electron-hole pairs) of the nearby metal
particles by a linear coupling between the dipole and a number of boson
modes. To be specific, the dipole points along the $z$ axis
and is placed at $z=h$ between two flat metal surfaces at $z=0$ and
$z=L$. The interaction Hamiltonian can then be written
\begin{equation}
   H_{int} = \sum_{\vec{q},\alpha} 
   (
      C_{\vec{q},\alpha} b_{\vec{q},\alpha}
      +
      C_{\vec{q},\alpha}^* b_{\vec{q},\alpha}^{\dag}
   )
   \hat{p}.
\end{equation}
Here $\hat{p}$ is the molecule dipole operator, $b$ and $b^{\dag}$ are
annihilation and creation operators for the bosons (with in-plane wave
vector $\vec{q}$ and another branch index $\alpha$) and the coupling
coefficients $C_{\vec{q},\alpha}$  are dependent on the position $h$ of
the dipole.  Using the Fermi golden rule, $H_{int}$  now gives a decay
rate from the excited state to the ground state of the molecule which
can be written
\begin{equation}
   w =
   \frac{2\pi}{\hbar}
   |p_0|^2
   A
   \int \frac{d^2q}{(2\pi)^2}
   \sum_{\alpha} 
   |C_{\vec{q},\alpha}|^2 
   \delta(\hbar\Omega_{ge} - \hbar\omega_{\vec{q},\alpha}).
\label{eq:ehrate1}
\end{equation}

The above expression is only useful if we have a way of calculating the
coefficients $C_{\vec{q},\alpha}$.  To do this we evaluate the energy
dissipation from a classical dipole placed at the position of the
molecule to the bosonic degrees of freedom.  Again, this expression will
contain the coefficients $C_{\vec{q},\alpha}$, but with a classical
dipole the energy dissipation rate can also be calculated using
classical electrodynamics, and expressed in terms of geometric
parameters and the dielectric function of the metal. 
We write the classical dipole moment as 
\begin{equation}
   p(t) = p_1(t) + p_1^*(t) =
   p_1 e^{-i\omega t} + p_1^* e^{i\omega t}.
\label{eq:classdipole}
\end{equation}
By letting $p(t)$ take the place of $\hat{p}$ in the Hamiltonian,
we get, from the Fermi golden rule, an energy dissipation rate 
\begin{equation}
   W =
   2 \pi \omega 
   |p_1|^2
   \sum_{\vec{q},\alpha}
   |C_{\vec{q},\alpha}|^2 
   \delta(\hbar\omega - \hbar\omega_{\vec{q},\alpha}).
\label{eq:Wgolden}
\end{equation}

Next we must calculate the dissipated power $W$ within the
framework of classical electrodynamics.  To this end we place the
dipole with a dipole moment  given by Eq.\ (\ref{eq:classdipole})
between the metal surfaces, at $z=h$.  Since all the distances involved
in this calculation are very short we can ignore effects of retardation
and express the solution in terms of a scalar potential
$\phi(t)=\phi_1e^{-i\omega t}+\phi_1^*e^{i\omega t}$, where $\phi_1$ can
be expressed as a Fourier transform 
\begin{equation}
   \phi_1(x,y,z) 
   =
   \int \frac{d^2q}{(2\pi)^2}
   \tilde{\phi}_q(z) 
   e^{i \vec{q}\cdot\vec{r}_{\|}},
\end{equation}
($\vec{r}_{\|}=x\hat{x}+y\hat{y}$).
For $\phi$ to satisfy Poisson's equation 
\begin{equation}
   \nabla^2 \phi = - \rho/\varepsilon_0, 
\end{equation}
(where $\rho$ is the charge density due to the dipole),
$\tilde{\phi}_q(z)$ must be a
linear combination of two exponentials $e^{q_{\|}z}$ and $e^{-q_{\|}z}$. 
In the classical calculation we
treat the surfaces in terms of their surface response functions:
$g_1(\vec{q},\omega)$ for the surface at $z=0$ and $g_2(\vec{q},\omega)$
for the surface at $z=L$. 
The surface
response function gives the negative ratio between the ``reflected''
potential (decaying as one leaves the surface) and the ``incident''
potential (decaying as one approaches the surface.  Thus,
\begin{equation}
   g(\vec{q},\omega) 
   = - 
   \tilde{\phi}_q^{\mathrm{refl}}
   /
   \tilde{\phi}_q^{\mathrm{inc}},
\end{equation}
where the potentials should be evaluated right at the surfaces.
Within d-parameter theory the surface response function is given to
first order in $q_{\|}=|\vec{q}|$ as\cite{Liebsch:87} 
\begin{equation}
   g(\vec{q}, \omega) 
   =
   \frac{\varepsilon(\omega)-1}{\varepsilon(\omega)+1}
   \,
   \left[
      1+
      \frac{\varepsilon(\omega)}{\varepsilon(\omega)+1}
      \,
      2 q_{\|} d_{\perp}(\omega)
   \right].
\label{eq:surfresp}
\end{equation}
The local dielectric function $\varepsilon(\omega)$ is the same as used in
the Mie calculation and $d_{\perp}(\omega)$ is the d-parameter function.
In our calculations we have evaluated $\Imag [d_{\perp}]$ from
Table~I of Ref.\ \onlinecite{Liebsch:87} using $r_s=3$ and
$\hbar\omega_p=9 \unit{eV}$ appropriate for silver in the low-frequency
regime. The real part $\Real [d_{\perp}]$ plays a less important role; we set
it to the constant value $\Real [d_{\perp}] = 1$ \r{A} here.

The solution for $\phi_1$ can now be
expressed in terms of either the total incident potential at $z=0$,
$\tilde{\phi}_{q,\mathrm{down}}$, as 
\begin{equation}
   \phi_1(x,y,z)
   =
   \int \frac{d^2q}{(2\pi)^2}
   e^{i\vec{q}\cdot\vec{x}}
   \tilde{\phi}_{q,\mathrm{down}} 
   \left( e^{qz} - g_1 e^{-qz} \right),
\label{eq:pot_at_down}
\end{equation}
or,  by a similar expression, in terms of
$\tilde{\phi}_{q,\mathrm{up}}$, the incident potential at the upper
interface. 
$\tilde{\phi}_{q,\mathrm{down}}$ and
$\tilde{\phi}_{q,\mathrm{up}}$ are given by
\begin{equation}
   \tilde{\phi}_{q,\mathrm{down}}
   =
   -
   \frac{p_1}{2\varepsilon_0} 
   \frac{e^{-qh} + g_2 e^{-q{(2L-h)}}}
   {1-g_1g_2 e^{-2qL}},
\end{equation}
and 
\begin{equation}
   \tilde{\phi}_{q,\mathrm{up}}
   =
   \frac{p_1}{2\varepsilon_0} 
   \frac{e^{-q(L-h)} + g_1 e^{-q{(L+h)}}}
   {1-g_1g_2 e^{-2qL}}.
\end{equation}

The Poynting vector $\vec{S}= \vec{E}\crossprod\vec{H}$ at the two
interfaces can here be approximated by an expression involving only
$\phi$; the dissipated power is then given by 
\begin{equation}
   W =
   4 \varepsilon_0
   \int \frac{d^2q}{(2\pi)^2}
   \omega q
   \left[
      |\tilde{\phi}_{q,\mathrm{down}}|^2
      \Imag g_1
      +
      |\tilde{\phi}_{q,\mathrm{up}}|^2
      \Imag g_2
   \right].
\label{eq:Wclassic}
\end{equation}
But this power should equal the one found in Eq.\ (\ref{eq:Wgolden}),
and in this way we get a relation between the classical quantities found
here and the sum over $\alpha$ in Eq.\ (\ref{eq:Wgolden}), (the sum over
$q$ goes over to an integral).  Inserting
the so obtained expression into Eq.\ (\ref{eq:ehrate1}) gives
\begin{widetext}
\begin{equation}
   w =
   \frac{2\pi}{\hbar} |p_0|^2
   \int_{q_{\mathrm{min}}}^{\infty} \frac{dq_{\|}}{(2\pi)^2}
   {\varepsilon_0 q_{\|}^2}
   \left[
      \left|\frac{2 \tilde{\phi}_{q,\mathrm{down}}}{p_1}\right|^2
      \Imag g_1
      +
      \left|\frac{2\tilde{\phi}_{q,\mathrm{up}}}{p_1}\right|^2
      \Imag g_2
   \right].
\label{eq:ehrate}
\end{equation}
\end{widetext}
The electron-hole contribution to $|M_d|^2$ in Eq.\ (\ref{eq:Mdtot}) can
now be found as 
\begin{equation}
   P_{eh}/P_{\mathrm{free}}
   =
   w/\gamma_{\mathrm{rad},0},
\end{equation}
where $\gamma_{\mathrm{rad},0}$ is found from Eq.\ (\ref{eq:gammarad0})
with $|f|^2=1$.

\begin{figure}[tb]
   \includegraphics[angle=0,width=8.0 cm]{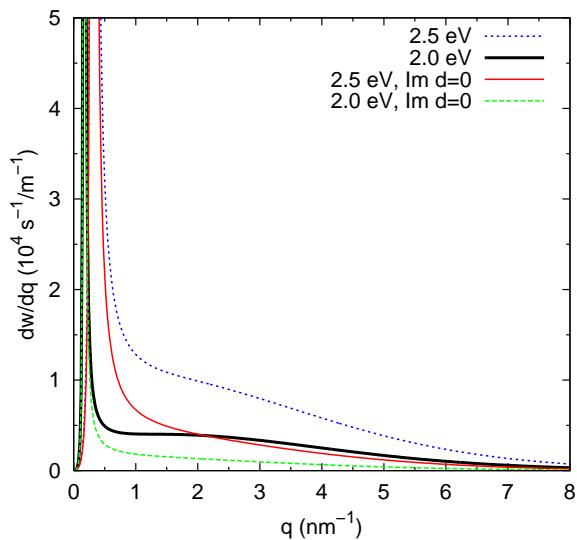}
   \caption{(color online)
      Plot of the function being integrated in Eq.\
      (\ref{eq:ehrate}). 
   }
   \label{fig:ehintegrand}
\end{figure}

The integrand of Eq.\ (\ref{eq:ehrate}) has been plotted, for two
different photon energies, as a function of $q$ in Fig.\
\ref{fig:ehintegrand}, both with and without a nonzero imaginary part
for the function $d_{\perp}(\omega)$.  This illustrates different
contributions to the damping rate.  At small $q$ we have damping
mediated by relatively long-wavelength interactions between the molecule
and the substrates. This part of the damping is actually captured by
the Mie calculation discussed in the main text.  In this range of $q$
space we see a fairly sharp peak around 0.5
$\unit{nm^{-1}}$ that is due to losses to a resonant interface plasmon
mode formed between the two metal surfaces. We also see that the use of
a non-local dielectric function, a nonzero value for $\mathrm{Im}
d_{\perp}(\omega)$, has essentially no effect here. Moving towards higher
$q$ values we encounter contributions to the damping that are not
included in the Mie calculation.  There are two reasons for this: (i) It
becomes technically difficult to go to very high values for the angular
momentum $\ell$ and consequently rapid (high $q$) variations of the
fields are not accounted for. (ii) At these larger wave vectors the
non-local effects ($\mathrm{Im} d_{\perp}(\omega)\neq0$)
that are not so easily included in the Mie calculation
provides the dominant contribution to the damping. This is clearly seen
in Fig.\ \ref{fig:ehintegrand}.

In order not to double-count contributions to the damping already
included in the Mie calculation we employ a low-$q$ cutoff in 
Eq.\ (\ref{eq:ehrate}) using $q_{\mathrm{min}}=1 \unit{nm^{-1}}$. It is
not really possibly to determine an exact position for the cutoff, since
there is no exact correspondence between angular momenta and wave
vectors, 
but 1 $\unit{nm^{-1}}$ is a reasonable value to use together
with $R=40 \unit{nm}$  and $\ell_{\mathrm{max}}=50$.

\end{document}